\documentclass{aa}
\usepackage{algorithm}
\usepackage[noend]{algpseudocode}

\usepackage{graphicx}
\usepackage{txfonts}
\bibpunct{(}{)}{;}{a}{}{,}

\begin{document}

   \title{\texttt{fast-resolve}: Fast Bayesian Radio Interferometric Imaging}

   \author{Jakob Roth
          \inst{1}\inst{2}\inst{3}\fnmsep\thanks{\email{roth@mpa-garching.mpg.de}}
          \and
          Philipp Frank
		  \inst{1}
		  \and
		  Hertzog L. Bester
		  \inst{4}\inst{5}
        \and
        Oleg M. Smirnov
        \inst{5}\inst{4}\inst{6}
        \and
        R\"udiger Westermann
        \inst{3}
		  \and
		  Torsten A. En{\ss}lin
		  \inst{1}\inst{2}
	  }

   \institute{Max Planck Institute for Astrophysics, Karl-Schwarzschild-Str. 1, 85748 Garching, Germany
         \and
             Ludwig-Maximilians-Universit\"at, Geschwister-Scholl-Platz 1, 80539 Munich, Germany
         \and
		 	Technische Universität M\"unchen (TUM), Boltzmannstr. 3, 85748 Garching, Germany
         \and
         South African Radio Astronomy Observatory (SARAO), Cape Town, 7925, South Africa
         \and
         Centre for Radio Astronomy Techniques and Technologies (RATT), Department of Physics and Electronics, Rhodes University, Makhanda, 6140, South Africa
         \and
         Institute for Radioastronomy, National Institute of Astrophysics (INAF IRA), Via Gobetti 101, 40129 Bologna, Italy
		 }

   \date{Received XXXX; accepted XXXX}

  \abstract
   {Interferometric imaging is algorithmically and computationally challenging as there is no unique inversion from the measurement data back to the sky maps, and the datasets can be very large. Many imaging methods already exist, but most of them focus either on the accuracy or the computational aspect.}
   {This paper aims to reduce the computational complexity of the Bayesian imaging algorithm \texttt{resolve}, enabling the application of Bayesian imaging for larger datasets.}
   {By combining computational shortcuts of the \texttt{CLEAN} algorithm with the Bayesian imaging algorithm \texttt{resolve} we developed an accurate and fast imaging algorithm which we name \texttt{fast-resolve}.}
   {We validate the accuracy of the presented \texttt{fast-resolve} algorithm by comparing it with results from \texttt{resolve} on VLA Cygnus~A data. Furthermore, we demonstrate the computational advantages of \texttt{fast-resolve} on a large MeerKAT ESO~137-006 dataset which is computationally out of reach for \texttt{resolve}.}
   {The presented algorithm is significantly faster than previous Bayesian imaging algorithms, broadening the applicability of Bayesian interferometric imaging. Specifically for the single channel VLA Cygnus~A datasets \texttt{fast-resolve} is about $144$ times faster than \texttt{resolve}. For the MeerKAT dataset with multiple channels the computational speedup of \texttt{fast-resolve} is even larger.}


   \maketitle

\section{Introduction}

Interferometric imaging is a versatile technique in astronomy that allows us to achieve enormous sensitivity and resolution by combining multiple telescopes. The effective resolution is roughly equivalent to that of a single telescope with a diameter equal to the largest distance between individual stations of the interferometer, which can be thousands of kilometers. For the upcoming Square Kilometer Array \citep{Labate22,Swart22}, the total collecting area might eventually approach one square kilometer, resulting in superior sensitivity compared to any single-dish telescope. Overcoming the resolution and sensitivity limitations of single telescopes comes at the cost of making it more difficult to retrieve images from the observational data.

Mathematically, the recovery of the sky images from interferometric data can be formulated as an inverse problem. The forward relation that computes the corresponding measurement data from a given sky image is known as the radio interferometric measurement equation \citep{Smirnov_2011}. As discussed in detail in Sec. \ref{sec:invprob}, the measured data points are essentially noisy and undersampled Fourier modes of the sky image. This makes direct inversion of the measurement equation to obtain the sky image impossible, turning radio interferometric imaging into an ill-posed inverse problem.

Solving the inverse problem of radio interferometric imaging requires sophisticated algorithms that impose additional constraints on the sky brightness and regularize possible solutions to it. Historically, CLEAN \citep{Hogbom74,Schwab83} has been by far the most widely used imaging algorithm because it is computationally efficient, simple, and easy to use.
Over the last decades, CLEAN-based imaging algorithms have been significantly improved, especially for diffuse emission imaging, spectral imaging, and wide-field imaging \citep{Bhatnagar04,Cornwell2008,Rau11,Offringa2014,Offringa2017}.
However, CLEAN-based algorithms have several drawbacks, such as limited image fidelity, suboptimal resolution of recovered images, and lack of uncertainty quantification (see eg. \cite{Arras2021} for a detailed discussion).
To improve these limitations, many other imaging algorithms have been developed.

A large class of new imaging algorithms builds on applying compressed sensing techniques to astronomical data \citep{Wiaux09}. Several incarnations of such algorithms have shown significant improvements in terms of image fidelity and resolution over CLEAN based reconstructions. Recent examples are \cite{Dabbech2018,Abdulaziz19,Dabbech_2021} utilizing sparsity based regularizers in combination with convex optimization. \cite{Birdi18, Birdi_2020} extended these approaches to full polarization imaging. In \cite{Repetti19} and \cite{Liaudat23}, some form of uncertainty quantification was added. \cite{Dabbech22,Thouvenin23} parallelized the image regularizer to distribute it over multiple CPUs.
\cite{Terris23,Dabbech22} presented a regularizer building on a neural network image denoisers. \cite{2023arXiv231207137T} further explored neural network regularizers and studied the influence of the training dataset on the regularizer.

Bayesian imaging algorithms are another important class of imaging algorithms addressing the uncertainty quantification problem. Early Bayesian imaging approaches such as \cite{Cornwell85} built on the maximum entropy principle. Other Bayesian imaging techniques such as \cite{Sutton06, Sutter14, Cai18, Tiede22} rely on posterior sampling techniques. The Bayesian imaging framework \texttt{resolve}\footnote{https://gitlab.mpcdf.mpg.de/ift/resolve} originally proposed by \cite{Junklewitz16} builds on variational inference instead of sampling based techniques for posterior approximation, reducing the computational costs. \texttt{resolve} has already successfully been applied to VLA, EHT, VLBA, GRAVITY, and ALMA data, providing high-quality radio maps with superior resolution compared to CLEAN, as well as an uncertainty map.
Recent examples are \cite{Arras_2019} joining Bayesian imaging with calibration. In \cite{Arras2021} \texttt{resolve} is compared with CLEAN, while in \cite{Roth23}, direction-dependent calibration is added and compared with sparsity based imaging results of \cite{Dabbech_2021}. In \cite{Stadler22} \texttt{resolve} is applied to the optical interferometer GRAVITY. In \cite{Arras2022} \texttt{resolve} is applied to EHT data.

The improvements in imaging algorithms mentioned above come at the cost of increased computational complexity.
For Very Long Baseline Interferometer (VLBI) observations, data sizes are typically small and the higher computational cost of advanced imaging methods is unproblematic.
However, for large arrays such as MeerKAT \citep{2016mks..confE...1J}, this limits the applicability of many of these algorithms.
For this reason, only very few advanced imaging algorithms have so far been successfully applied to data sets the size of a typical MeerKAT observation.
\cite{Dabbech22} presents an application of an imaging algorithm using sparsity based and neural network regularizers to the MeerKAT ESO~137-006 observation.
To handle the enormous computational complexity, \cite{Dabbech22} massively parallelize their imaging algorithm and distribute it across a high performance computing system, requiring hundreds to thousands of CPU hours to converge. In \cite{Thouvenin23}, this approach of parallelizing the regularizer is extended to the spectral domain.

To the best of our knowledge, no Bayesian radio interferometric imaging algorithm has so far been applied to a similar sized dataset. The goal of the \texttt{resolve} framework has always been to enable Bayesian imaging for a wide range of radio telescopes, not just VLBI observations.
However, for instruments such as MeerKAT, imaging with \texttt{resolve} becomes computationally prohibitively expensive.
In this paper, we present the algorithm \texttt{fast-resolve}, which significantly reduces the computational complexity of classical \texttt{resolve} and enables Bayesian image reconstruction for large data sets.
\cite{greiner16} previously already named an imaging method fast\texttt{RESOLVE}.
The goal of the fast\texttt{RESOLVE} algorithm of \cite{greiner16} is the same as the new incarnation of \texttt{fast-resolve} presented in this paper.
However, although the new and the old variants of \texttt{fast-resolve} share some ideas, there are significant differences.
In Sec. \ref{sec:fastrveold}, we will highlight the common concepts between the old the and the new \texttt{fast-resolve}.
Note that the framework behind both \texttt{fast-resolve} and \texttt{resolve} has evolved significantly since the time of \cite{greiner16}.

Machine learning-based imaging algorithms are a third relatively new class of imaging algorithms. Once the underlying machine learning models are trained, the computational cost for imaging is often lower than for other algorithms.
Nevertheless, assessing image fidelity is challenging in such a framework given the lack of interpretability of machine learning models.
Recent examples of such algorithms are \cite{Aghabiglou24,Aghabiglou23,Connor22,Schmidt22}.

The remainder of the paper is organized as follows. In Sec. \ref{sec:invprob}, we discuss the radio interferometric measurement equation and the imaging inverse problem in detail. In Sec. \ref{sec:method}, we briefly review the existing \texttt{resolve} framework in its current form and outline the \texttt{CLEAN} algorithm. Building on some of the computational shortcuts of the \texttt{CLEAN} algorithm, we derive \texttt{fast-resolve}, and in Sec. \ref{sec:apply}, we show several applications of it. In particular, in Sec. \ref{sec:applycomp} we compare imaging results of \texttt{fast-resolve} with the classical \texttt{resolve} framework on VLA \citep{perly11} data to validate the fidelity of the resulting sky images, and in Sec. \ref{sec:applyspeed} we present a \texttt{fast-resolve} reconstruction of the ESO~137-006 MeerKAT observation to demonstrate the computational speedup.

\section{The inverse problem}\label{sec:invprob}
The radio interferometric measurement equation, derived, for example, in \cite{Smirnov_2011}, relates the radio sky brightness to the datapoints, often called visibilities. More specifically, via the measurement equation, model visibilities $\tilde{V}$ can be computed for an assumed sky brightness $I$ and antenna sensitivity $G$.
Under the assumption of scalar antenna gains $G$, the  model visibility $\tilde{V}_{pqt}$ of antennas $p$ and $q$ at time $t$ is given as:
\begin{align}\label{eq:rime}
   \tilde{V}_{pqt} = \int C(\boldsymbol{l},w_{pqt})I(\boldsymbol{l})G_{p}(t,\boldsymbol{l})G^{*}_{q}(t,\boldsymbol{l}) e^{-2\pi i (\boldsymbol{k}_{pqt}\cdot\boldsymbol{l})} d\boldsymbol{l},
\end{align}
where
\begin{itemize}
   \item $\boldsymbol{l} = (l,m)$ are the sky coordinates,
   \item $t$ is the time coordinate,
   \item $\boldsymbol{k}_{pqt} = (u_{pqt}, v_{pqt})$ are the baseline uv-coordinates in units of the imaging wavelength,
   \item $C(\boldsymbol{l},w_{pqt}) = \exp{\left(-2\pi i w_{pqt}\left(\sqrt{1-\boldsymbol{l}^2} - 1\right)\right)} / \sqrt{1-\boldsymbol{l}^2}$ is the $w$- or non-coplanar baselines effect,
   \item $I(\boldsymbol{l})$ is the sky brightness,
   \item $G_{p}(t,\boldsymbol{l})$ is the gain of antenna $p$ depending on time and potentially also direction.
\end{itemize}
The model visibilities are therefore Fourier-like components of the sky brightness $I$ modulated by the antenna gains $G_p$ and $G_q$ as well as the $w$-effect $C$. The visibilities actually recorded in the measurement are related to the model visibilities according to
\begin{align}\label{eq:rs}
   V_{pqt} = \tilde{V}_{pqt} + n = R(I) + n,
\end{align}
with $n$ representing some unknown noise in the measurement, and $R$ the mapping from sky brightness to visibilities defined in Eq. \ref{eq:rime}. With Eq. \ref{eq:rime} and Eq. \ref{eq:rs}, it is straightforward to compute simulated visibilities for an assumed sky brightness $I$, antenna gain $G$, and noise statistics. Inverting this relation is not possible without additional assumptions, since first, the measured visibilities $V$ are corrupted by the measurement noise $n$, and second,  Eq. \ref{eq:rime} is generally not invertible since in practical applications not all Fourier components are measured.

The non-uniqueness of solutions to Eq. \ref{eq:rs} means that additional regularisation needs to be imposed to discriminate between all possible sky images compatible with the data.
This additional information is either an explicit prior or regularization term, or is implicitly encoded in the structure of the imaging algorithm.

\section{Methods}\label{sec:method}
In this section, we derive the \texttt{fast-resolve} algorithm (Sec. \ref{sec:fastrveintro}). Since \texttt{fast-resolve} builds on the already existing Bayesian imaging framework \texttt{resolve}, we start with a brief review of the classic \texttt{resolve} imaging method (Sec. \ref{sec:rveintro}). As some of the computational speedups of \texttt{fast-resolve} are inspired by the \texttt{CLEAN} imaging algorithm, we also outline the basic concepts behind \texttt{CLEAN} (Sec. \ref{sec:cleaninto}).

\subsection{\texttt{resolve}} \label{sec:rveintro}

The \texttt{resolve} imaging algorithm addresses the imaging inverse problem from a probabilistic perspective. Thus, instead of reconstructing a single estimate of the sky brightness, it infers the posterior probability distribution $P(I|V)$ of possible sky images given the measured visibilities. Bayes' Theorem
\begin{align}
    P(I|V) = \frac{P(V|I)P(I)}{P(V)}
\end{align}
expresses the posterior probability $P(I|V)$ in terms of the likelihood $P(V|I)$, the prior $P(I)$, and the evidence $P(V)$.
\texttt{resolve} provides models for the likelihood $P(V|I)$ and the prior $P(I)$. The posterior distribution can be inferred for a given prior and likelihood model building on the functionality of the Bayesian inference package NIFTy\footnote{https://github.com/NIFTy-PPL/NIFTy} \citep{nifty1,nifty3,niftyre}. In the next three subsections we briefly outline the likelihood and prior of \texttt{resolve} as well as the variational inference algorithms of NIFTy.

\subsubsection{\texttt{resolve} prior} \label{sec:rvepr}
The \texttt{resolve} framework provides predefined prior models for the two types of radio emission, point sources and extended diffuse emission. Both priors encode that the brightness must be positive, since there is no negative flux. Furthermore, both priors are very flexible and allow for brightness variations over several orders of magnitude.

For the point source prior, pixels are independently modeled with an inverse gamma prior for their intensity. The inverse gamma distribution is strictly positive and has a wide tail, allowing extremely bright sources. In the example in Sec. \ref{sec:applycomp}, such a prior is applied for the two bright point sources in the core of Cygnus~A.
While the brightness of point sources is reconstructed from the data, the locations currently need to be manually set. This limits the applicability of the current point source prior, as discussed in the application to MeerKAT ESO~137-006 data (Sec. \ref{sec:applyspeed}).

Besides positivity and possible variations over several orders of magnitude, the diffuse emission prior also encodes correlations of the brightness of nearby pixels, which is essentially the defining property of diffuse emission. The correlation of nearby pixels in the diffuse emission prior is modeled by Gaussian processes. The result of the Gaussian process is exponentiated to ensure positivity. Detailed explanations of the \texttt{resolve} prior models can be found in \cite{Arras2021}. An important aspect of all the prior models in \texttt{resolve} is that they are fast and scalable to very large numbers of pixels. For example, in \cite{Arras2021} a $4096\times2048$ pixel Gaussian process-based diffuse emission prior is used to image Cygnus~A at various frequencies.

\subsubsection{\texttt{resolve} likelihood} \label{sec:rvelh}

The likelihood is evaluated in \texttt{resolve} using Eq. \ref{eq:rime}. More specifically, using Eq. \ref{eq:rime}, we can write the likelihood as $P(V|I) = P(V|R(I)) = P(V|\tilde{V})$. The noise statistics in Eq. \ref{eq:rs} then determines the likelihood. In \texttt{resolve}, Gaussian noise statistics are assumed. For numerical reasons, \texttt{resolve} works with the negative logarithm of the likelihood, named the likelihood Hamiltonian, instead of the likelihood itself. With the Gaussian noise assumption, the likelihood Hamiltonian is given by
\begin{align}\label{eq:ham}
    H(V|I) = \frac{1}{2}\left(V - R(I) \right)^{\dagger} N^{-1} \left(V - R(I) \right) + \frac{1}{2}\ln |2\pi N|,
\end{align}
with $\dagger$ denoting complex conjugate transpose and $\frac{1}{2}\ln |2\pi N|$ coming form the normalization of the Gaussian, which can be ignored in many applications.
Different possibilities are implemented for the covariance of the noise $N$. In the simplest case, the weights of the visibilities are used as the inverse noise covariance. Alternatively, the noise covariance can also be estimated during the image reconstruction.

The computationally important aspect of the likelihood in the classic \texttt{resolve} framework is that for every update step where the likelihood is evaluated, also $R(I)$ and thus Eq. \ref{eq:rime} need to be computed. This can be computationally expensive, especially for datasets with many visibilities, as we will discuss in detail later.

\subsubsection{\texttt{resolve} posterior inference}
\texttt{resolve} is built on the probabilistic programming package NIFTy, which provides variational inference methods \citep{Knollmueller2019, Frank2021} to approximate the posterior distribution for a given prior and likelihood function. The advantage of variational inference techniques over sampling techniques such as MCMC or HMC is that they scale better with the number of parameters, which is the number of pixels in the imaging context. For example, the variational inference method of NIFTy has recently been used in a 3D reconstruction with 607 million voxels \citep{Edenhofer23}. While the variational inference algorithms scale very well with the number of parameters, they still need to evaluate the likelihood very often.
This means variational inference is fast as long as the evaluation of the likelihood and prior is fast.

As discussed in Sec. \ref{sec:rvelh}, evaluating the likelihood in \texttt{resolve} boils down to evaluating the radio interferometric measurement equation (Eq. \ref{eq:rime}). To do so \texttt{resolve} relies on the parallelizable \texttt{wgridder} \citep{Arras2021_gr} implemented in the \texttt{ducc}\footnote{https://gitlab.mpcdf.mpg.de/mtr/ducc} library. Nevertheless, evaluating the measurement equation becomes computationally intensive for data sets with many visibilities. For example, for the MeerKAT data set considered in Sec. \ref{sec:applyspeed}, evaluating Eq. \ref{eq:rime} on eight threads takes about $45$ seconds. This becomes prohibitive for algorithms which require many thousands of likelihood evaluations.

\subsection{\texttt{CLEAN}} \label{sec:cleaninto}
In this section, we briefly outline some of the concepts behind the \texttt{CLEAN} algorithm \citep{Hogbom74,Schwab83}, as the computational speedups of \texttt{fast-resolve} over classic \texttt{resolve} are partly inspired by \texttt{CLEAN}.
We will refrain from delving into the details behind the numerous improvements that have been made to the \texttt{CLEAN} algorithm over the last decades (see \cite{Rau09} for example) and focus only on the aspects which are relevant to speeding up \texttt{resolve}. We refer the reader to \cite{Arras2021} for a detailed comparison between \texttt{CLEAN} and \texttt{resolve}.

\texttt{CLEAN} is an iterative optimization algorithm minimizing the weighted square residuals between the measured visibilities and the model visibilities computed from the sky brightness model. Expressed as a formula, the objective function minimized by \texttt{CLEAN} is identical to Eq.\ref{eq:ham}, the likelihood Hamiltonian of \texttt{resolve}. Minimizing Eq. \ref{eq:ham} with respect to the sky brightness $I$ is equivalent to solving
\begin{align}\label{eq:clean}
    R^\dagger N^{-1} R I = R^\dagger N^{-1} V,
\end{align}
with $R^{\dagger}$ being the adjoint operation of $R$, thus mapping from visibilities to the sky brightness.
Neglecting wide-field effects originating from non-coplanar baselines, the operation $R^\dagger N^{-1} R$ is, equivalent to a convolution with the effective point spread function of the interferometer $I^{\text{PSF}}$. $R^\dagger N^{-1} V$ is the back projection of the noise-weighted data into the sky domain, called dirty image $I^{\texttt{D}}$. This can be expressed with the formula
\begin{align}
    I^{\text{PSF}} * I \approx R^\dagger N^{-1} V = I^\text{D},
\end{align}
with $I^{\text{PSF}}$ being the PSF of the interferometer and $*$ denoting convolution.
Thus, the dirty image $I^{\text{D}} = R^\dagger N^{-1} V$ is approximately the true sky brightness $I$ convolved with the PSF $I^{\text{PSF}}$ of the interferometer. Radio interferometric imaging is, therefore, nearly equivalent to deconvolving the dirty image.

As discussed in Sec. \ref{sec:invprob}, no unique solution to imaging inverse problems exists. The absence of a unique solution manifests as $R^\dagger N^{-1} R$ and $I^{\text{PSF}} *$ not being invertible operations. While in \texttt{resolve}, additional regularization was provided via an explicit prior, regularization of the sky images in \texttt{CLEAN} is implicitly encoded into the structure of the algorithm. More specifically, \texttt{CLEAN} starts with an empty sky model as an initial estimate $I^{m}_{0}$ for the true sky brightness and iteratively adds components, in the simplest form point sources, to this estimate until a stopping criterion is met. This introduces the implicit prior that the sky is sparsely represented by a finite number of \texttt{CLEAN} components.

In the following, we briefly summarize the algorithmic structure of \texttt{CLEAN} as relevant for \texttt{fast-resolve}. The iterative procedure of adding components to the current model image $I^{m}$ is split into major and minor cycles. In the major cycles, a current residual image $I^{\text{RES}} = R^\dagger N^{-1} (V - R I^{m}_{i})$ is computed, with $I^{m}_{i}$ being the current model image. In the minor cycle, additional components are added to the model, and their PSFs are removed from the residual image. In the subsequent major cycle, the residual image is recomputed for the updated model image, and in the following new minor cycle, more components get added to the model. This scheme is iterated until a global stopping criterion is met.

From a computational perspective, the important aspect of the major/minor scheme is that $R$ and $R^\dagger$ only need to be evaluated in major cycles. In minor cycles, the PSF of the added \texttt{CLEAN} components is subtracted from the residual image, but for this, no evaluations of $R$ and $R^\dagger$ are needed since the PSF can be precomputed once. Most of the computations of the \texttt{CLEAN} algorithm are performed in image space, and only major cycles go back to the visibility space. In contrast, \texttt{resolve} needs to map from image to data space by applying $R$ for every likelihood evaluation.

\subsection{\texttt{fast-resolve}} \label{sec:fastrveintro}

As outlined in the previous section, evaluating the radio interferometric instrument response (Eq. \ref{eq:rime}) contributes a substantial fraction to the overall runtime of a \texttt{resolve} image reconstruction. Therefore, reducing the number of necessary evaluations of Eq. \ref{eq:rime} has the potential for significant speedups. The basic idea of \texttt{fast-resolve} is to perform most of the computations in image space, similar to \texttt{CLEAN}, and only evaluate the radio interferometric instrument response once in a while.

\subsubsection{\texttt{fast-resolve} measurement equation}
The radio interferometric measurement equation in the form of Eq. \ref{eq:rs} leads to the likelihood Hamiltonian Eq. \ref{eq:ham} of the classic \texttt{resolve} framework, which involves an evaluation of the interferometric instrument response $R$.
To get a likelihood Hamiltonian that does not include evaluating $R$, one has to transform the measurement equation such that all involved quantities live in image space, not data space.
To project all involved quantities to image space, we apply $R^\dagger N^{-1}$ from the left to Eq. \ref{eq:rs}, and get:
\begin{align}
    R^\dagger N^{-1} V &= R^\dagger N^{-1} R I + R^\dagger N^{-1} n\\
    I^\text{D} &= R' I + n',\label{eq:rs_new}
\end{align}
with $R' = R^\dagger N^{-1} R$, and $n' = R^\dagger N^{-1} n$.
The quantities of the new measurement equation are $I^\text{D}$, $I$, and $n'$ and are all defined in image space and not in data space.
The statistics of the new noise $n'$ remains Gaussian as $R^\dagger N^{-1}$ is a linear transformation. The corresponding likelihood Hamiltonian is given by:
\begin{align}\label{eq:ham_new}
    H(I^\text{D}|I) = \frac{1}{2}(I^\text{D} - R' I)^\dagger N'^{-1}(I^\text{D} - R' I)  + \frac{1}{2}\ln |2\pi N'|.
\end{align}
The covariance of the transformed noise is given by
\begin{align}
    N' &= \left< n'n'^\dagger \right> = R^\dagger N^{-1} \left<n n^\dagger \right> N^{-1} R =  R^\dagger N^{-1} R.
\end{align}
This new noise covariance $N'$ is not invertible as $R^\dagger N^{-1} R$ is not invertible, which is problematic as the likelihood Hamiltonian contains the inverse noise covariance $N'^{-1}$. To mitigate the problem of a singular noise covariance, we modify the measurement equation once more by adding uncorrelated Gaussian noise with a small amplitude. This leads to a noise covariance
\begin{align}\label{eq:N_new}
    N' = R^\dagger N^{-1} R + \epsilon \mathbb{I},
\end{align}
with $\mathbb{I}$ being the unit matrix and $\epsilon$ a small number, which is the variance of the additional noise. With this artificially introduced additional noise, the full noise covariance $N'$ is invertible, and the new likelihood Hamiltonian Eq. \ref{eq:ham_new} becomes well defined.

\subsubsection{\texttt{fast-resolve} response $R'$} \label{sec:fastrver}
How large the speedup of using the new likelihood Hamiltonian Eq. \ref{eq:ham_new} is compared to the old Hamiltonian Eq. \ref{eq:ham} depends on how much faster the new Hamiltonian can be evaluated. The idea of \texttt{fast-resolve} is to approximate $R' = R^\dagger N^{-1} R $ with a PSF convolution $R' = R^\dagger N^{-1} R \approx I^{\text{PSF}}*$, as it is done in a similar way in the \texttt{CLEAN} algorithm. Approximating $R' = R^\dagger N^{-1} R $ by a convolution with the PSF is only exact for coplanar arrays. Corrections for the inaccuracy of the approximation are applied in some major cycles in analogy to the \texttt{CLEAN} algorithm, as we describe in Sec. \ref{sec:fastrvei}.

The convolution with the PSF $I^{\text{PSF}}*$ can be efficiently applied via an FFT convolution. For data sets with many visibilities, this FFT-based convolution with the PSF $I^{\text{PSF}}$ has the potential to be significantly faster than an evaluation of the interferometer response $R$ (Eq. \ref{eq:rime}). In Sec. \ref{sec:apply}, we will compare the computational speedup of the new likelihood for different data sets.

As the sky brightness $I$ and the PSF $I^{\text{PSF}}$ are non-periodic, some padding of the sky is needed for an FFT-based convolution.
More specifically, to exactly evaluate $I^{\text{PSF}}*I$, the PSF with which we convolve needs to be twice as big as the field of view we want to image since some emission in the sky $I$ could be at the edge of the field we are imaging. The sky image needs to be padded with zeros to the same size as the PSF. As a formula, this can be noted as
\begin{align}
    R'I &= R^\dagger N^{-1} R I\\
        & \approx I^{\text{PSF}}* I \\
        & = P^{\dagger}\text{FFT}^{-1}\left[\text{FFT}[I^{\text{PSF}}] \cdot \text{FFT}[PI] \right],
\end{align}
with $P$ denoting the padding operation, and $P^{\dagger}$ for slicing out the region not padded. By neglecting PSF sidelobes and reducing its size, the necessary amount of zero padding can be reduced.
This, however, reduces the accuracy of the approximation $R' \approx I^{\text{PSF}}*$, which might make it necessary to perform more major cycles in the image reconstruction.

\subsubsection{\texttt{fast-resolve} noise model} \label{sec:fastrven}
To evaluate the likelihood Hamiltonian of \texttt{fast-resolve}, we need to apply both $R'= R^\dagger N^{-1} R $ and $N'^{-1}= (R^\dagger N^{-1} R + \epsilon \mathbb{I})^{-1}$. Similar to $R'$, we need to approximate $N'^{-1}$ such that we can apply it without having to evaluate $R$ and $R^\dagger$ every time. The basic idea of the approximation of $N'^{-1}$ is the same as for $R'$, thus replacing the exact operation with an FFT convolution. Expressed as a formula, we compute the application of $N'$ to some input $x$ via
\begin{align}
    N'(x) \approx \text{FFT}^{-1} \left[ K \cdot\text{FFT}(x) \right],
\end{align}
with some appropriate Kernel $K$. The approximate noise covariance still needs to fulfill the mathematical properties of covariances, because for our Bayesian model, the likelihood is a probability and not just an arbitrary cost function. Therefore, the approximated noise covariance matrix needs to be Hermitian and positive definite. Specifically, for the convolution approximation, this implies that all entries of the convolution kernel $K$ must be real and strictly positive.
To fulfill these constraints, we parameterize $K$ as
\begin{align}
    K_{\xi} = \exp \xi + \epsilon,
\end{align}
with $\epsilon$ as defined in Eq. \ref{eq:N_new} and $\xi$ being an implicitly defined real-valued vector.
We set this vector $\xi$ such that it minimizes the square residual between the true and the approximate noise covariance when applied to a test image with a point source in the center. Thus $\xi$ is set to
\begin{align}
    \bar{\xi} = \text{argmin}_\xi\left(N'(I_{\delta}) - \text{FFT}^{-1} \left[ (\exp \xi + \epsilon) \cdot\text{FFT}(I_{\delta}) \right] \right)^2
\end{align}
with $I_{\delta}$ having a point source or delta peak in the center of the field of view and otherwise being zeros.
In words, we approximate the noise covariance $N'$ with an appropriate Kernel $K$ that yields a proper covariance by construction,
is easy to invert, and minimizes the squared distance to the effect that $N'$ has on a point source.

To evaluate the likelihood (Eq. \ref{eq:ham_new}) we need to apply $N'^{-1}$. We do this by convolving with the inverse kernel:
\begin{align}
    N'^{-1}(x) \approx \text{FFT}^{-1} \left[ K_{\bar{\xi}}^{-1} \cdot\text{FFT}(x) \right].
\end{align}
This involves a second approximation, as we here implicitly assume periodic boundary conditions. As long as the main lobe of the PSF is much smaller than the field of view, this assumption does not create large errors.

\subsubsection{\texttt{fast-resolve} inference scheme} \label{sec:fastrvei}

The previous sections derived the approximate likelihood for \texttt{fast-resolve} inspired by the \texttt{CLEAN} algorithm.
In the same spirit, we adapt the major/minor cycle scheme of \texttt{CLEAN} to utilize it with Bayesian inference for a probabilistic sky brightness reconstruction. In the minor cycles, we optimize the current estimate of the posterior distribution $P(I|V)$ for the sky brightness $I$ using the above approximations for a fast likelihood evaluation. In the major cycles, we apply, similar to \texttt{CLEAN}, the exact response operations to correct for the approximation error.

The algorithm starts by initializing the dirty image $I^{\text{D}} = R^{\dagger}N^{-1}V$ from the visibilities $V$ using the exact response function $R$. The dirty image is the input data $d_0 = I^{\text{D}}$ for the first minor cycle. The first minor cycle computes an initial estimate of the posterior distribution $P_0$ of the sky brightness using the measurement equation
\begin{align}
    d_0 = I^{\text{PSF}}*I + n',
\end{align}
with the approximations discussed in Sec. \ref{sec:fastrver} and Sec. \ref{sec:fastrven}. We will discuss the exact algorithm used to estimate the posterior distribution in Sec. \ref{sec:fastrveminor}. From this initial estimate of the posterior distribution $P_0$, we compute the posterior mean of the sky brightness $I_0 = \left<I\right>_{P_0}$. The posterior mean $I_0$ is the output of the first minor cycle and the input to the first major cycle. The first major cycle computes the residual $d_1$ between the dirty image $I^{\text{D}}$ and the posterior mean of the first minor cycle passed through the response $R^{\dagger}N^{-1}R$ of the fast resolve measurement equation:
\begin{align}
    d_1 = I^{\text{D}} - R^{\dagger}N^{-1}RI_0.
\end{align}
In contrast to the minor cycle, $R^{\dagger}N^{-1}R$ is here computed exactly and not approximated with $I^{\text{PSF}}*$. The residual $d_1$ is the output of the major cycle and the input to the second minor cycle. In the second minor cycle, the posterior $P_1$ of $I$ for the measurement equation
\begin{align}
    d_1 = I^{\text{PSF}}*(I - I_0) + n',
\end{align}
is approximated. This approximation can be done efficiently by starting with the posterior estimate of the previous minor cycle and refining this estimate. The output of the second minor cycle is the updated posterior mean $I_1 = \left<I\right>_{P_1}$ which is the input to the next major cycle computing the new residual to the dirty image using the exact response:
\begin{align}
    d_2 = I^{\text{D}} - R^{\dagger}N^{-1}RI_1.
\end{align}
The new residual data $d_2$ is the input to the next minor cycle. This scheme is iterated until converged, thus until no significant structures are left in the residual $d_n$ and the posterior estimates $I_{n}$ are not changing anymore. In algorithm \ref{alg:fast-resolve} the full inference scheme is summarized as a pseudocode algorithm.

\begin{figure}[]
    \begin{algorithm}[H]
      \caption{\texttt{fast-resolve} inference scheme}\label{alg:fast-resolve}
      \begin{algorithmic}[0]
        \State \# compute dirty image
        \State $I^\text{D} \gets R^{\dagger} N^{-1} V$
        \State
        \State \# initialize input with dirty image
        \State $d_0 \gets I^\text{D}$
        \State
        \State \# reconstruct $n$ major cycles
        \For{i in  $0\ldots n - 1$}
            \State \# minor cycle
            \State estimate $P_{i}(I)$ from $d_i = I^{\text{PSF}}*I + n'$
            \State \# compute posterior mean
            \State $I_i \gets \left<I\right>_{P_i}$
            \State
            \State \# major cycle updating input data
            \State $d_{i+1} \gets I^{\text{D}} - R^{\dagger}N^{-1}RI_i$
        \EndFor
        \State
        \State \# final result: $P_{n - 1}(I)$
      \end{algorithmic}
    \end{algorithm}
\end{figure}

The overall structure of the \texttt{fast-resolve} major/minor scheme is very similar to the major/minor scheme of \texttt{CLEAN}. The main difference is that \texttt{fast-resolve} updates in the minor cycles a probability distribution for possible sky images instead of a simple model image, as is the case for \texttt{CLEAN}. The algorithm for inferring the posterior distribution is outlined in the following subsection \ref{sec:fastrveminor}.

\subsubsection{\texttt{fast-resolve} minor cycles} \label{sec:fastrveminor}
In the minor cycles, we optimize the approximation of the posterior distribution $P(I|V)$ using the scalable variational inference algorithms \citep{Knollmueller2019, Frank2021} of the NIFTy package already employed in the classic \texttt{resolve} algorithm. These variational inference algorithms account for correlations between parameters of the model. While the method Metric Gaussian Variational Inference (MGVI) of \cite{Knollmueller2019} relies on a Gaussian approximation of the posterior distribution, the algorithm geometric Variational Inference (geoVI) of \cite{Frank2021} can also capture non-Gaussian posterior distributions. Although \texttt{fast-resolve} relies for the minor cycles on the same inference algorithms as \texttt{resolve}, updating the posterior distribution is much faster in \texttt{fast-resolve} since the approximate likelihood described above is used instead of evaluating the exact measurement equation.

Furthermore, the NIFTy package has also undergone a major rewrite, switching from a NumPy based \citep{NumPy-Array} implementation to a JAX\footnote{https://github.com/google/jax} \citep{jax2018github} based backend, allowing for GPU-accelerated computing. While \texttt{resolve} was so far building on the old NumPy-based NIFTy, \texttt{fast-resolve} makes use of the JAX accelerated NIFTy version, also named NIFTy.re \citep{niftyre}.
Especially the FFT convolutions in \texttt{fast-resolve} have the potential for significant GPU acceleration.
In Sec. \ref{sec:apply}, we compare the runtime of \texttt{resolve} and \texttt{fast-resolve} using the CPU and the GPU backend.

In the future, we also plan to port the classic \texttt{resolve} framework to JAX. Porting \texttt{resolve} is more involved than porting \texttt{fast-resolve} because it requires binding a high performance implementation of the radio interferometric measurement operator to JAX. As a preparatory work, we have developed the JAXbind\footnote{https://github.com/NIFTy-PPL/JAXbind} package \citep{Roth24} which allows to bind custom functions to JAX. However, since the \texttt{wgridder} from ducc\footnote{https://gitlab.mpcdf.mpg.de/mtr/ducc} is used to evaluate the radio interferometric measurement equation, the JAX version of \texttt{resolve} will also be limited to run on the CPU.

\subsection{Previous fast\texttt{RESOLVE} of \cite{greiner16}} \label{sec:fastrveold}
The algorithm named fast\texttt{RESOLVE} introduced by \cite{greiner16} was an earlier attempt to speed up \texttt{resolve}. Similar to the \texttt{fast-resolve} of this work, \cite{greiner16} used approximations of the likelihood avoiding applications of $R$ in every step. More specifically, \cite{greiner16} derived a maximum a posteriori estimator of the sky brightness using the gridded weights of the visibilities as a noise covariance. Since fast\texttt{RESOLVE} had some limitations, such as it could not account for the w-term and could not provide uncertainty estimates, this approach was not followed up in subsequent developments of \texttt{resolve} in \cite{Arras_2019,Arras2021}, or \cite{Roth23}.

\section{Applications}\label{sec:apply}
For verification, we demonstrate the proposed \texttt{fast-resolve} method on different data sets. First, we reconstruct the Cygnus~A VLA observation at four different frequency bands, and second, we image ESO~137-006 using a MeerKAT observation. The Cygnus~A data is suitable to validate the accuracy as the \texttt{fast-resolve} images can be compared to previous results from \cite{Arras2021} and \cite{Roth23} obtained on the same dataset.
The Cygnus~A reconstruction has a relatively small field of view, and wide field effects should be negligible.
The MeerKAT observation is significantly larger than the VLA observation, and imaging with \texttt{resolve} is computationally out of reach, allowing to demonstrate the computational advantages of \texttt{fast-resolve} over \texttt{resolve}.
Furthermore, the field of view of the ESO~137-006 observations is also significantly larger such that the major cycles of \texttt{fast-resolve} correcting for wide field effects become more important.

\subsection{Application to VLA Cygnus~A data}\label{sec:applycomp}
\subsubsection{Data and algorithm setup} \label{sec:applycygaintro}
For imaging Cygnus~A, we use the exact same data already imaged in \cite{Arras2021} with \texttt{resolve} and \texttt{CLEAN}. The data contains single frequency channels at $2052\,\text{MHz}$ (S-band), $4811\,\text{MHz}$ (C-band), $8427\,\text{MHz}$ (X-band), and $13360\,\text{MHz}$ (Ku-band). Also, in \cite{Roth23}, an uncalibrated version of the S-band data was jointly calibrated and imaged with \texttt{resolve}. In all observing bands, all four VLA configurations were used. Computationally relevant details of the observations, such as the number of visibilities and the grid size of the reconstructed images, are listed in Tab. \ref{tab:obscyg}. For details on the calibration of the data, we refer to \cite{Sebokolodi20}.

\begin{table}
    \caption{VLA Cygnus~A observations. All observations are single channel and used all four VLA configurations. For all frequencies the same field of view (Fov) is imaged.}
    \label{tab:obscyg}
    \centering
    \begin{tabular}{c c c c}
    \hline\hline
    Freq [MHz]& $N_{rows}$ & image size & Fov [deg]\\
    \hline
    2052  & 1281930 & $512\times1024$   & $0.025\times0.05$ \\
    4811  & 1281934 & $1024\times2048$  & $0.025\times0.05$ \\
    8427  & 2523954 & $2048\times4096$  & $0.025\times0.05$ \\
    13360 & 2544102 & $2048\times4096$  & $0.025\times0.05$ \\
    \hline
    \end{tabular}
\end{table}

As a prior model, we use the combination of the diffuse emission prior and the point source prior described in Sec. \ref{sec:rvepr}. This prior model is identical to the prior model already used in the existing \texttt{resolve} framework. Specifically, in \cite{Arras2021} and \cite{Roth23}, this prior model was already used for Cygnus~A reconstructions.
For completeness, we summarize the main aspects of the prior model here.
We refer to \cite{Arras2021} for details on the prior model.
We model the diffuse emission with an exponentiated Gaussian Process spanning the entire field of view. For the two point sources in the nucleus of Cygnus~A, we insert two separate point source models at the locations of these sources. For the brightness of these sources we use an inverse gamma prior. The exact hyper-parameters for the Gaussian process and the inverse gamma prior are listed in appendix \ref{app:paramcyga}.
The number of pixels used to model the diffuse emission is listed in Tab. \ref{tab:obscyg} for the different frequencies.

In \cite{Arras2021}, the MGVI algorithm was used for inferring the posterior of the sky brightness map. For direct comparability between the results of this work and the previous \texttt{resolve} maps, we also use the MGVI in the minor cycles (Sec. \ref{sec:fastrveminor}) for the posterior approximation.
For the imaging of the S-band data in \cite{Roth23}, the geoVI algorithm was used. However, we do not expect significant differences in the resulting sky maps.

To summarize, \cite{Arras2021} and \cite{Roth23} used the same prior model setup as we use for imaging Cygnus~A with \texttt{fast-resolve}.
Furthermore, the posterior inference algorithm is expected to produce similar results for a given likelihood.
Therefore, the previous \texttt{resolve} results of \cite{Arras2021} and \cite{Roth23} are ideally suited to validate the \texttt{fast-resovle} likelihood approximation in a direct comparison on a dataset with negligible wide field effects.
Furthermore, we compare with the multi-scale \texttt{CLEAN} reconstruction of the S-band data from \cite{Arras2021}.
For a more detailed comparison with \texttt{CLEAN}, we refer to \cite{Arras2021}, where \texttt{resolve} was extensively compared to single and multi-scale \texttt{CLEAN} for all frequency bands.

\subsubsection{Comparison with previous \texttt{resolve} results}

\begin{figure*}
    \centering
       \includegraphics[width=17cm]{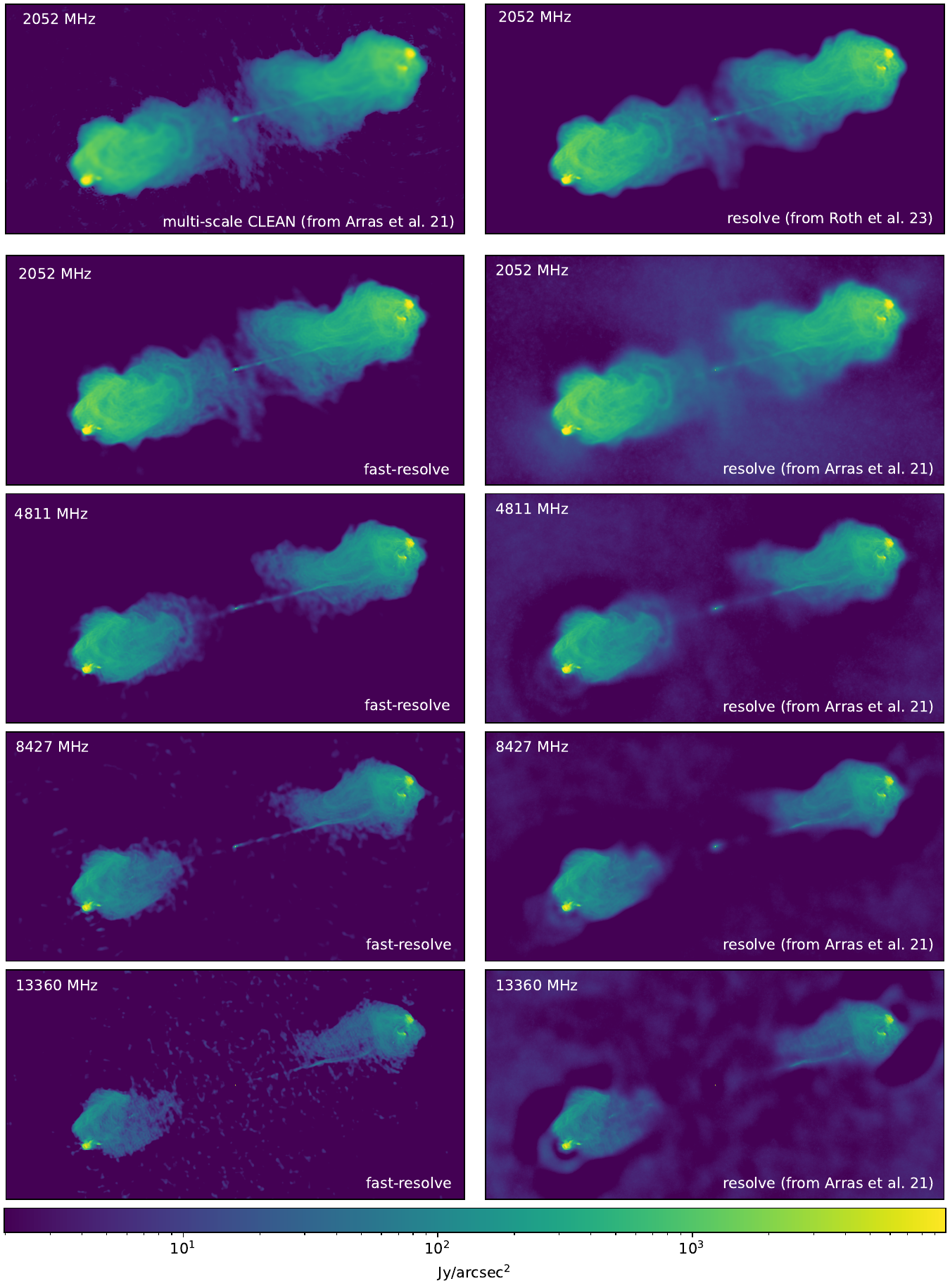}
         \caption{Comparison of \texttt{fast-resolve} Cygnus~A reconstructions with \texttt{resolve} and \texttt{CLEAN} reconstructions.
         The top left panel shows the multi-scale \texttt{CLEAN} map of the S-band data form \cite{Arras2021}.
         The left column below the \texttt{CLEAN} map shows \texttt{fast-resolve} reconstructions at four different frequencies indicated in the top left corner of each panel. The right column shows \texttt{resolve} reconstructions of \cite{Roth23} and \cite{Arras2021} using the same data. The dynamic range of the \texttt{fast-resolve} reconstructions is higher than in some of the previous \texttt{resolve} reconstructions. This is because the old \texttt{resolve} reconstructions were not fully converged due to their high computational cost.}
         \label{fig:fast_VS_slow}
 \end{figure*}

In the following, we compare the \texttt{fast-resolve} reconstructions with previous results to validate the accuracy of \texttt{fast-resolve}.
Fig. \ref{fig:fast_VS_slow} displays the \texttt{resolve} reconstructions of \cite{Arras2021} and \cite{Roth23} as well as the S-band data multi-scale \texttt{CLEAN} reconstruction in comparison with the \texttt{fast-resolve} reconstructions of this work.
The overall quality of the \texttt{fast-resolve} maps is on par with the \texttt{resolve} results.
The multi-scale \texttt{CLEAN} reconstruction has a lower resolution in bright regions of the lobes than the \texttt{resolve} based reconstructions.
All bright emission features are consistently reconstructed by all algorithms in all frequency bands.
The \texttt{fast-resolve} and \texttt{CLEAN} maps have a higher dynamic range than the results of \cite{Arras2021}.
Nevertheless, this is not due to a conceptional problem of the \texttt{resolve} algorithm, but rather the reconstructions of \cite{Arras2021} were not fully converged due to the very high computational cost of \texttt{resolve}.
\texttt{resolve} reconstructs the brightest emission of the field before modeling fainter features.
Thus, faint features are missing in the radio map when a reconstruction is stopped before convergence.
For \texttt{fast-resolve}, where the overall runtime of the algorithm is much shorter, it is easier to ensure that the reconstruction is fully converged.
In Sec. \ref{sec:fastrve_time} the convergence of \texttt{resolve} and \texttt{fast-resolve} is analyzed in detail.
The \texttt{resolve} reconstruction of \cite{Roth23} has a similar dynamic range compared to the \texttt{fast-resolve} reconstructions.

\begin{figure*}
    \centering
       \includegraphics[width=17cm]{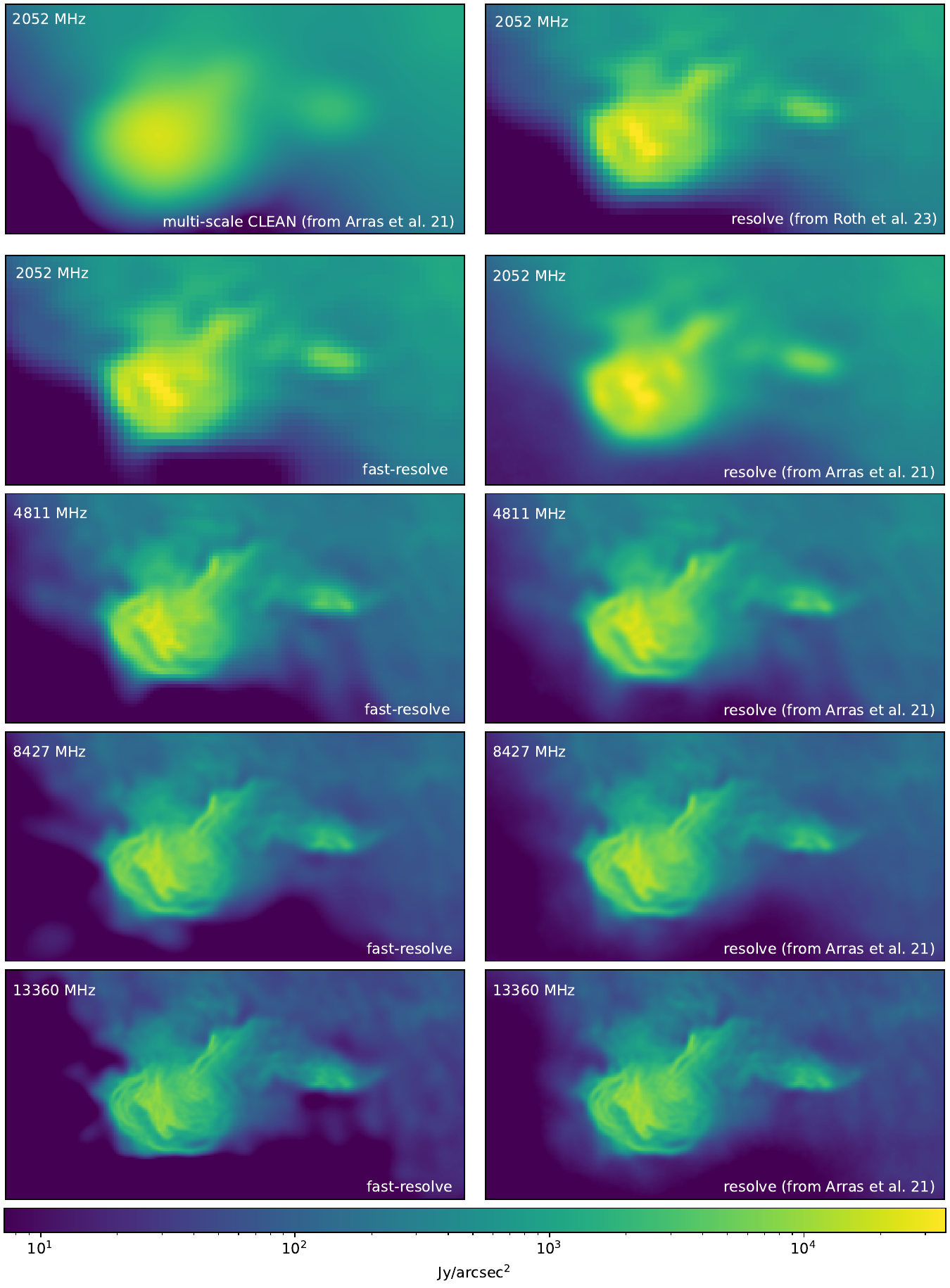}
         \caption{Zoom on the eastern hot spot of the Cygnus~A reconstructions from Fig. \ref{fig:fast_VS_slow}. The resolution of the \texttt{fast-resolve} maps in the left column is on par with the \texttt{resolve} maps in the right column. The resolution of the \texttt{CLEAN} map is significantly lower than the in the other maps.}
         \label{fig:fast_VS_slow_zoom_hotspot}
 \end{figure*}

In the comparison between \texttt{CLEAN} and \texttt{resolve} in \cite{Arras2021}, \texttt{resolve} produced significantly higher resolved maps than \texttt{CLEAN} in regions with high surface brightness.
A zoom into such a region, the eastern hotspot of Cygnus~A, is depicted in Fig. \ref{fig:fast_VS_slow_zoom_hotspot}. In this region, the results are consistent, and \texttt{fast-resolve} achieves the same resolution as the \texttt{resolve} based reconstructions.\footnote{The pixelizations of the multi-scale \texttt{CLEAN}, the previous \texttt{resolve}, and the \texttt{fast-resolve} reconstructions differ as they originate from different works.}
Furthermore, \texttt{fast-resolve} also shows minimal imaging artifacts around the hotspot, as does \texttt{resolve}.
The multi-scale \texttt{CLEAN} map from \cite{Arras2021} is added again for comparison. As already discussed in \cite{Arras2021}, the \texttt{CLEAN} reconstruction has a significantly lower resolution than the \texttt{resolve} reconstruction. These super-resolution capabilities where validated by comparing the morphological features of the \texttt{resolve} reconstructions with higher frequency observations under the assumption of spectral smoothness.

\begin{figure*}
    \centering
       \includegraphics[width=17cm]{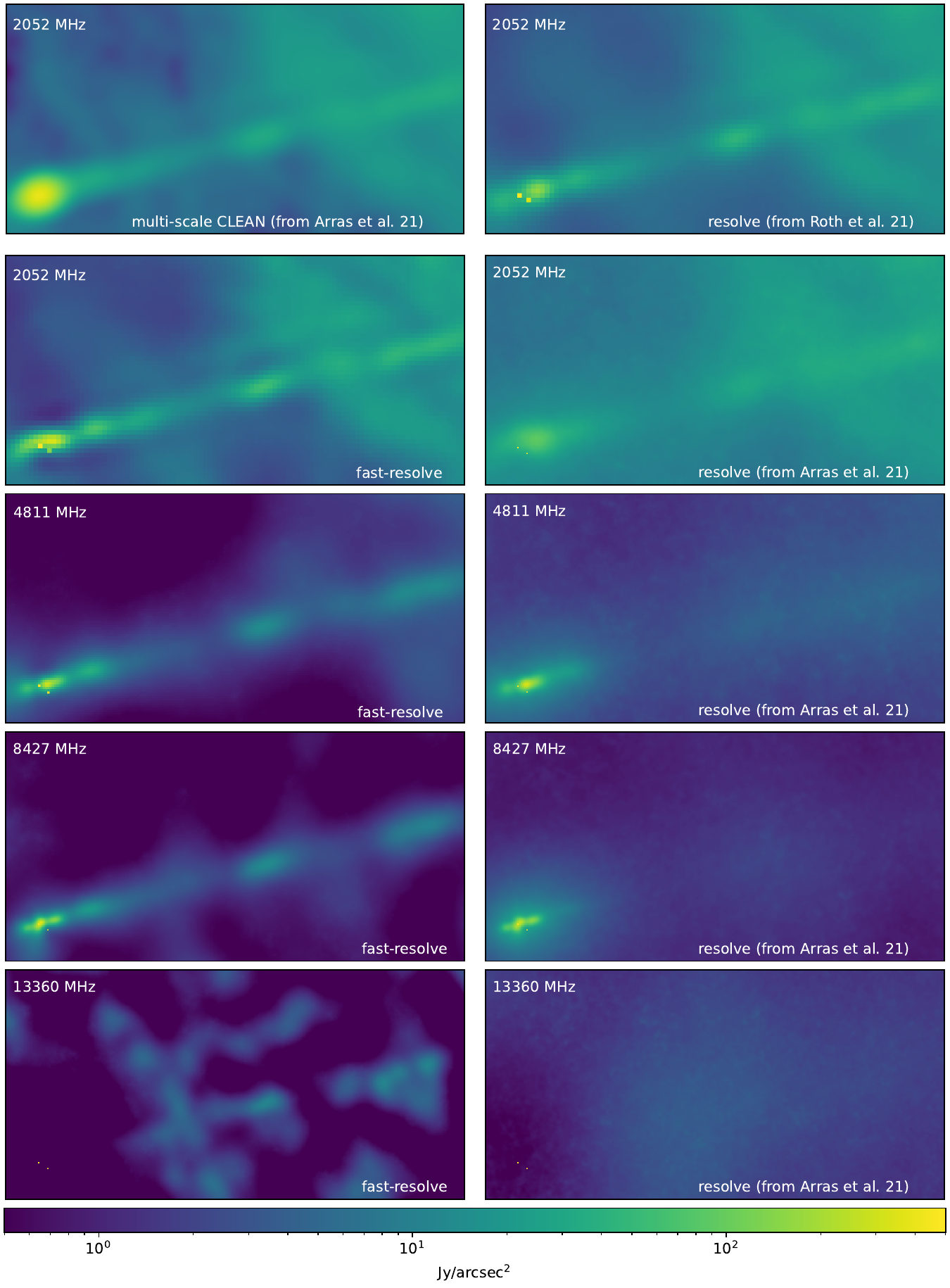}
         \caption{Zoom on the nucleus and jet of the Cygnus~A reconstructions form Fig. \ref{fig:fast_VS_slow}. In all \texttt{resolve} and \texttt{fast-resolve} reconstructions, the two point sources in the nucleus are modeled with two pixels that are uncorrelated with the brightness of the neighboring pixels. Since the \texttt{resolve} of \cite{Arras2021} are not fully converged due to their high computational cost, their dynamic range is lower, and the jet of Cygnus~A is not visible. The multi-scale \texttt{CLEAN} map from \cite{Arras2021} is consistent with the \texttt{fast-resolve} result and the \texttt{resolve} reconstruction from \cite{Roth23}. At the highest frequency the jet is hardly visible also with \texttt{fast-resolve}, with background artifacts having a similar brightness.}
         \label{fig:fast_VS_slow_zoom_jet}
 \end{figure*}

Fig. \ref{fig:fast_VS_slow_zoom_jet} zooms on the nucleus and jet of Cygnus~A.
In the zooms on the nucleus the pixels modeling the two point sources are clearly visible.
The \texttt{fast-resolve} S-band reconstruction and the \texttt{resolve} reconstruction of \cite{Roth23} show consistent results depicting the nucleus and jet. In the S-band reconstruction of \cite{Arras2021}, the jet is also visible, although due to the smaller dynamic range of the map, it is less pronounced. In the higher frequency bands, the shape of the core of Cygnus~A is consistent between the reconstructions of \texttt{resolve} and \texttt{fast-resolve}. As the old \texttt{resolve} reconstructions are not fully converged, their dynamic range is lower, and the fainter emission of the jet is barely visible at higher frequencies.
At the highest frequencies $13360\,\text{MHz}$, the jet is also barely visible in the \texttt{fast-resolve} reconstruction.
The \texttt{CLEAN} image of the jet is consistent with the \texttt{fast-resolve} reconstruction and the \texttt{resolve} reconstruction from \cite{Roth23}.

\begin{figure*}
    \centering
       \includegraphics[width=17cm]{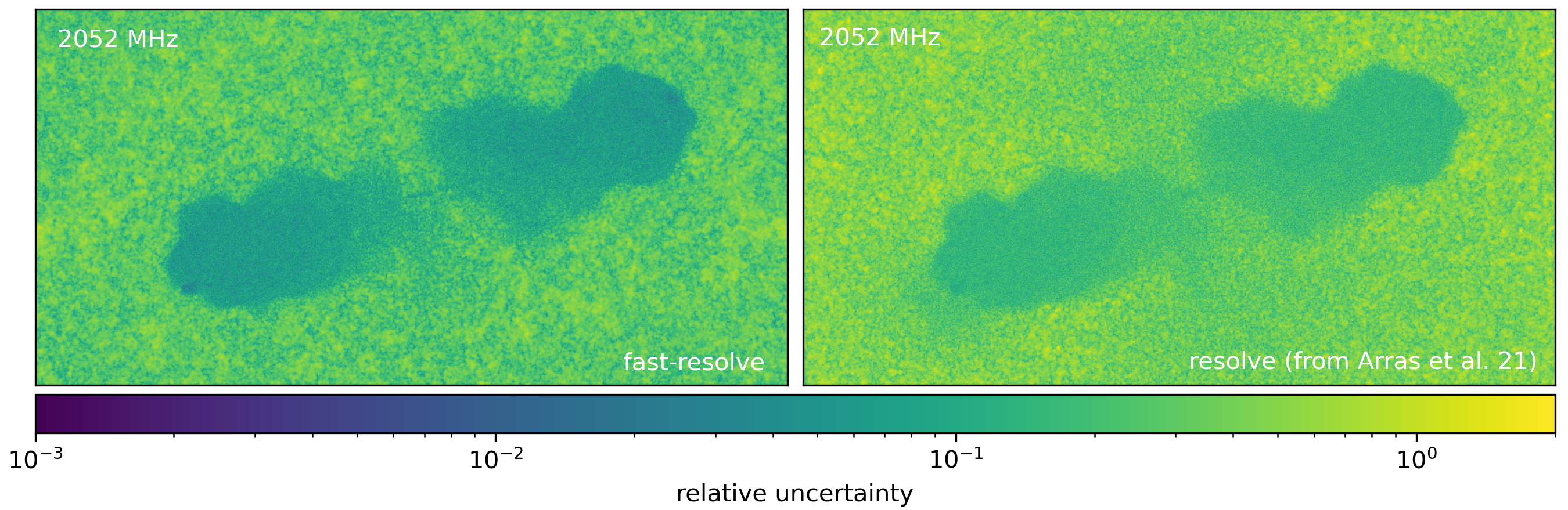}
         \caption{Pixel-wise relative uncertainty map of \texttt{fast-resolve} S-band reconstruction compared to the \texttt{resolve} uncertainty map of \cite{Arras2021}. The relative uncertainty is lower for both reconstructions in regions of high surface brightness. In the hotspot, the relative uncertainty is around $10^{-3}$ to $10^{-2}$ and increases towards the lower surface brightness regions. Outside of Cygnus~A, the relative uncertainty fluctuates around $1$, indicating that the reconstructed intensity is statistically consistent with zero surface brightness. The estimated uncertainty of \texttt{resolve} tends to be slightly higher than the uncertainty estimate of \texttt{fast-resolve}. This might be due to the \texttt{resolve} reconstruction of \cite{Arras2021} not being fully converged.}
         \label{fig:fast_VS_slow_std}
 \end{figure*}

\texttt{resolve} as well as \texttt{fast-resolve} provide posterior samples of the sky brightness distribution. From these samples, not only the posterior mean but also other summary statistics can be computed. As an example, we show in  Fig. \ref{fig:fast_VS_slow_std} the pixel-wise relative uncertainty of the S-band data reconstruction of \texttt{fast-resolve} in comparison with the corresponding uncertainty map of \cite{Arras2021}. The estimated uncertainty of \texttt{resolve} tends to be slightly higher than the uncertainty estimate of \texttt{fast-resolve}. This difference might come from the fact that the \texttt{resolve} reconstruction of \cite{Arras2021} was not fully converged since the uncertainty estimate of \texttt{resolve} usually becomes smaller when the algorithm converges. Nevertheless, we cannot exclude that this is related to the approximations of \texttt{fast-resolve}.

\subsubsection{Computational time of \texttt{fast-resolve}} \label{sec:fastrve_time}

In the previous subsection, we compared the results of \texttt{fast-resolve} with \texttt{resolve}, confirming that \texttt{fast-resolve} can deliver the same high quality radio maps as \texttt{resolve}. In this subsection, we want to analyze the computational speed of \texttt{fast-resolve} and compare it to the existing \texttt{resolve} framework.
To have a direct comparison between the runtimes, we imaged the S-band data with exactly the same hyperparameters for the prior model with \texttt{fast-resolve} and \texttt{resolve} and saved snapshots of the reconstructions at several points during the optimization. While \texttt{resolve} can only be executed on a CPU, \texttt{fast-resolve} can also run on a GPU.

\begin{figure*}
    \centering
       \includegraphics[width=17cm]{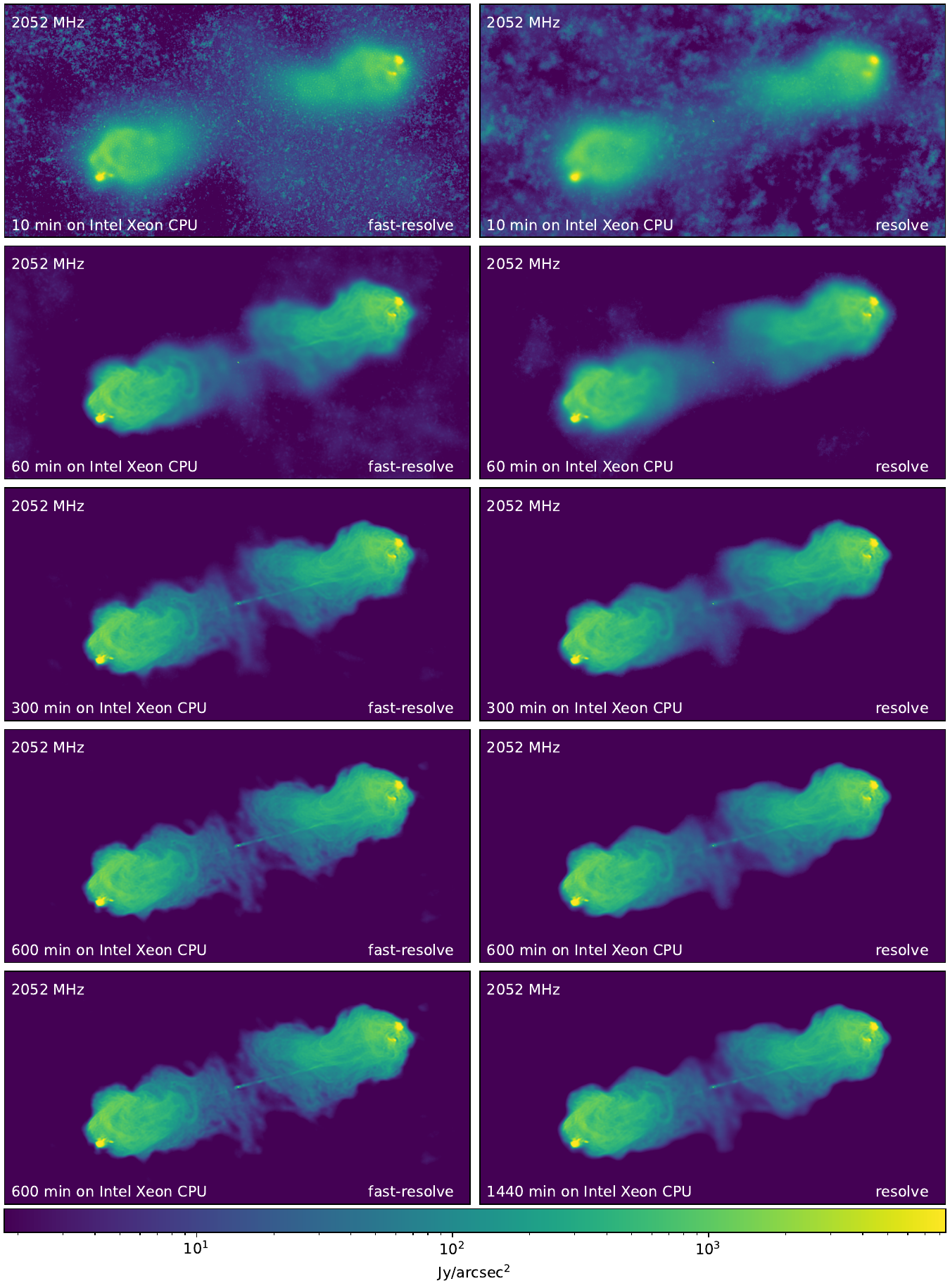}
         \caption{\texttt{fast-resolve} and \texttt{resolve} reconstructions of the $2052\,\text{MHz}$ Cygnus~A data at different stages of the reconstruction. The left column shows \texttt{fast-resolve} the right column \texttt{resolve}. The rows display snapshots of the reconstruction after different amounts of wall time indicated in the lower left of each panel. The \texttt{fast-resolve} reconstruction was performed on two cores of an Intel Xeon CPU. The \texttt{resolve} reconstruction used eight cores of the same CPU. The \texttt{fast-resolve} reconstructions in the last two rows are at identical time snapshots, since the \texttt{fast-resolve} reconstruction is considered to be converged after $600\,\text{min}$. The \texttt{resolve} reconstruction is not fully converged  even after $1416\,\text{min}$.}
         \label{fig:fast_VS_slow_zoom_time}
 \end{figure*}

Fig. \ref{fig:fast_VS_slow_zoom_time} depicts the direct comparison between \texttt{resolve} and \texttt{fast-resolve}, with both algorithms being executed on the same CPU. Specifically, the algorithms were run on an Intel Xeon W-1270P CPU with $3.80\,\text{GHz}$ clock speed and $8$ cores.
For the \texttt{resolve} reconstruction, the MGVI algorithm inferring the posterior distribution was parallelized with $4$ MPI tasks, each using $2$ cores for evaluating the radio interferometric measurement equation (Eq. \ref{eq:rs}), thus utilizing $8$ cores in total.
The \texttt{fast-resolve} reconstruction ran on the single python process without manual threading, relying on JAX to parallelize the computations over the cores.
Snapshots of the reconstructions after $10$, $60$, $300$, $600$, and $1440\,\text{min}$ are displayed. After $10\,\text{min}$, the hotspot and parts of the lobes are visible in the reconstructions of both algorithms. Nevertheless, the core and the jet are missing, and both reconstructions have significant background artifacts. After $60\,\text{min}$ runtime, fainter parts of the lobes become visible, and the \texttt{fast-resolve} map shows significantly more details in the lower surface brightness regions. After $300\,\text{min}$ the \texttt{fast-resolve} map also depicts the very low surface brightness regions with high resolution, while in the \texttt{resolve} map, the outflow is still partially missing, and the lower surface brightness regions of the lobes are reconstructed with low resolution. After $600\,\text{min}$ the \texttt{fast-resolve} map has only slightly changed compared to the $300\,\text{min}$ snapshot. At this point, we assume the \texttt{fast-resolve} algorithm to be fully converged. The \texttt{resolve} reconstruction is also after $600\,\text{min}$ not yet fully converged. The final snapshot of the \texttt{resolve} reconstruction is after $1440\,\text{min}$. At this stage also the \texttt{resolve} reconstruction is nearly converged. Only in the very low surface brightness regions of the lobes, the resolution is still lower than in the \texttt{fast-resolve} reconstruction.

\begin{figure*}
    \centering
       \includegraphics[width=17cm]{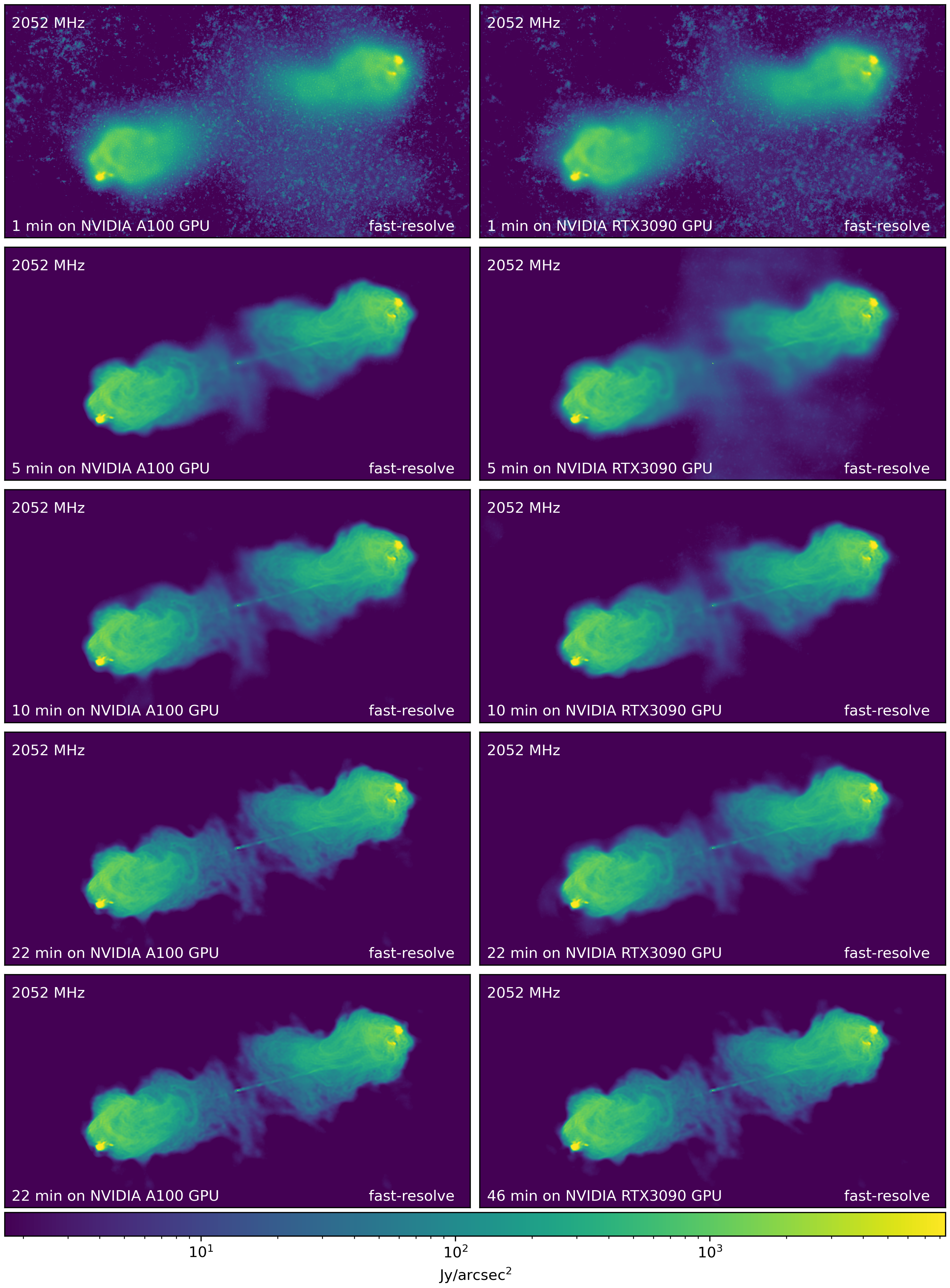}
         \caption{\texttt{fast-resolve} reconstructions of the $2052\,\text{MHz}$ Cygnus~A data. The left column displays snapshots of a \texttt{fast-resolve} reconstruction on an NVIDIA A100 GPU. The right column shows snapshots of a \texttt{fast-resolve} reconstruction on an NVIDIA RTX 390 GPU. Both reconstructions ran the same number of major and minor iterations. Since the reconstruction on the A100 GPU was finished after $22\,\text{min}$, the bottom left panel is identical to the row above. On the A100 GPU \texttt{fast-resolve} is approximately twice as fast as on a RTX 390 GPU, and 25 times faster than on the CPU displayed in Fig. \ref{fig:fast_VS_slow_zoom_time}.}
         \label{fig:fast_gpu_a_VS_r2}
 \end{figure*}

While Fig. \ref{fig:fast_VS_slow_zoom_time} analyzes the convergence of \texttt{fast-resolve} on the CPU, \texttt{fast-resolve} is mainly designed for GPUs. The performance of \texttt{fast-resolve} on the S-band data using GPUs is shown in Fig. \ref{fig:fast_gpu_a_VS_r2}. Specifically, Fig. \ref{fig:fast_gpu_a_VS_r2} displays snapshots of the \texttt{fast-resolve} reconstruction after $1$, $5$, $10$, $22$, and $46\,\text{min}$ using an NVIDIA A100 high-performance computing GPU compared with a reconstruction on a consumer-level GPU, an NVIDIA GeForce RTX 3090. When comparing the results, one should consider that the A100 GPU is on the order of 10 times more expensive than the RTX 3090. On both GPUs, we executed the exact same \texttt{fast-resolve} reconstruction we ran on a CPU in Fig. \ref{fig:fast_VS_slow_zoom_time}. Similar to the CPU run, the high surface brightness regions are reconstructed first before the method picks up the low surface brightness flux. On both GPUs, the algorithm is much faster than on the CPU. While on the CPU, the reconstruction was finished after $600\,\text{min}$, the same number of major and minor cycles were finished on the  RTX 3090 GPU in $46\,\text{min}$ and on the A100 GPU in $22\,\text{min}$. Thus, the \texttt{fast-resolve} reconstruction was 13 times faster on the RTX 3090 GPU and 27 times faster on the A100 GPU than the CPU reconstruction. In comparison, the \texttt{resolve} reconstruction on the CPU was not fully converged even after $1416\,\text{min}$.

\begin{figure*}
    \centering
       \includegraphics[width=17cm]{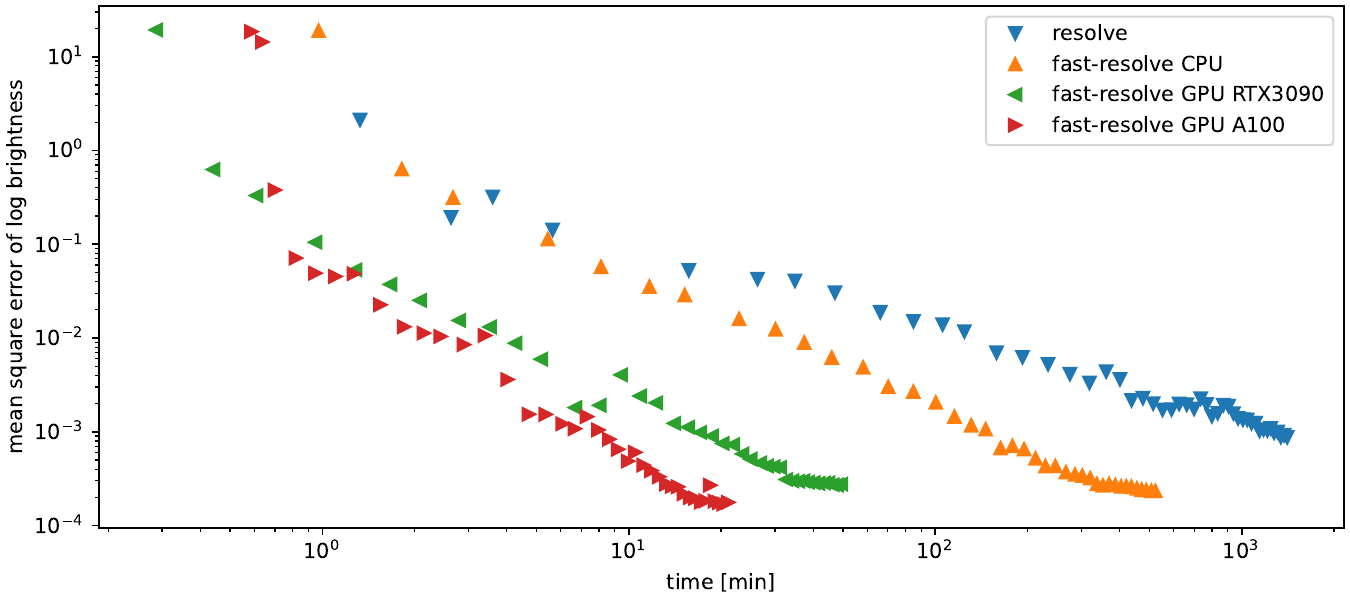}
         \caption{Mean square residual of the log brightness of \texttt{resolve} and \texttt{fast-resolve} reconstruction as a function of time for the S-band data. The residual is computed to the final iteration of \texttt{fast-resolve} on the NVIDIA A100 GPU depicted in Fig. \ref{fig:fast_gpu_a_VS_r2}, which we believe to be converged.
         The \texttt{fast-resolve} reconstructions are recomputed with a different random seed.
         The mean square residual of the \texttt{fast-resolve} reconstructions falls until their curves flatten when they converge. Due to the different hardware, JAX, and CUDA versions, the final results are numerically not exactly identical. The \texttt{resolve} reconstruction (blue curve) does not fully converge in the displayed time range.}
         \label{fig:convergence}
 \end{figure*}

To quantify the convergence rate of the reconstructions, we computed the mean square residual between the logarithmic brightness of the final \texttt{fast-resolve} iteration on the NVIDIA A100 GPU and earlier iterations of \texttt{resolve} and \texttt{fast-resolve} reconstructions.
We reran the \texttt{fast-resolve} reconstructions with different random seeds to be independent of the random seed used for the NVIDIA A100 reconstruction with respect to which the residuals are computed.
Fig. \ref{fig:convergence} shows the mean square residuals for \texttt{resolve} and the three \texttt{fast-resolve} reconstructions as a function of wall time.
The \texttt{fast-resolve} reconstructions converge within the displayed time, and their curves of the mean square residual flatten.
The reconstruction's final mean square residuals are not exactly the same since the final maps are not numerically identical because of the different hardware, JAX, and CUDA versions. The \texttt{resolve} reconstruction does not converge in the displayed time interval and the residual error keeps falling until the end after $1416\,\text{min}$ ($1$ day).
Of course, using the residual of the logarithmic brightness is an arbitrary metric for quantifying convergence speed. Nevertheless, it can roughly quantify the speedups of \texttt{fast-resolve}.
After $1416\,\text{min}$, the mean square residual of the \texttt{resolve} reconstruction is around $10^{-3}$.
In comparison, the \texttt{fast-resolve} reconstructions reach the same mean squared residual after around $10\,\text{min}$, $20\,\text{min}$, and $200\,\text{min}$ for runs on the NVIDIA A100 GPU, the NVIDIA RTX 3090 GPU and the Intel Xeon CPU, respectively.
Thus, the indicative speedup of \texttt{fast-resolve} over \texttt{resolve} on the Cygnus~A S-band data is a factor of $144$ for the A100 GPU, a factor of $72$ for the RTX 3090 GPU and a factor of $7.2$ for the CPU run.
The timings reported above do not include the time needed to precompute the convolution kernels for the response and noise of \texttt{fast-resolve}. Nevertheless, these are small compared to the time needed for imaging. For the S-band data, for example, the computation of the kernels takes only $0.6\,\text{min}$, which is much shorter than the imaging runtime, even on the GPU.

As indicated in Tab. \ref{tab:obscyg}, the Cygnus~A data has only a single frequency channel. For datasets with more baselines and frequency channels, such as the MeerKAT datasets considered in the next section (see Tab. \ref{tab:obseso}), the algorithmic advantage of \texttt{fast-resolve} of not having to compute the radio response in each evaluation of the likelihood is much larger.
Thus, for such datasets, the speedup of \texttt{fast-resolve} will be significantly larger than for the Cygnus~A single channel imaging.
Indeed for the MeerKAT dataset of the next section (Sec. \ref{sec:applyspeed}) imaging with \texttt{resolve} is computationally out of scope, while with \texttt{fast-resolve} images can still be reconstructed with moderate computational costs.

\begin{figure*}
    \centering
       \includegraphics[width=17cm]{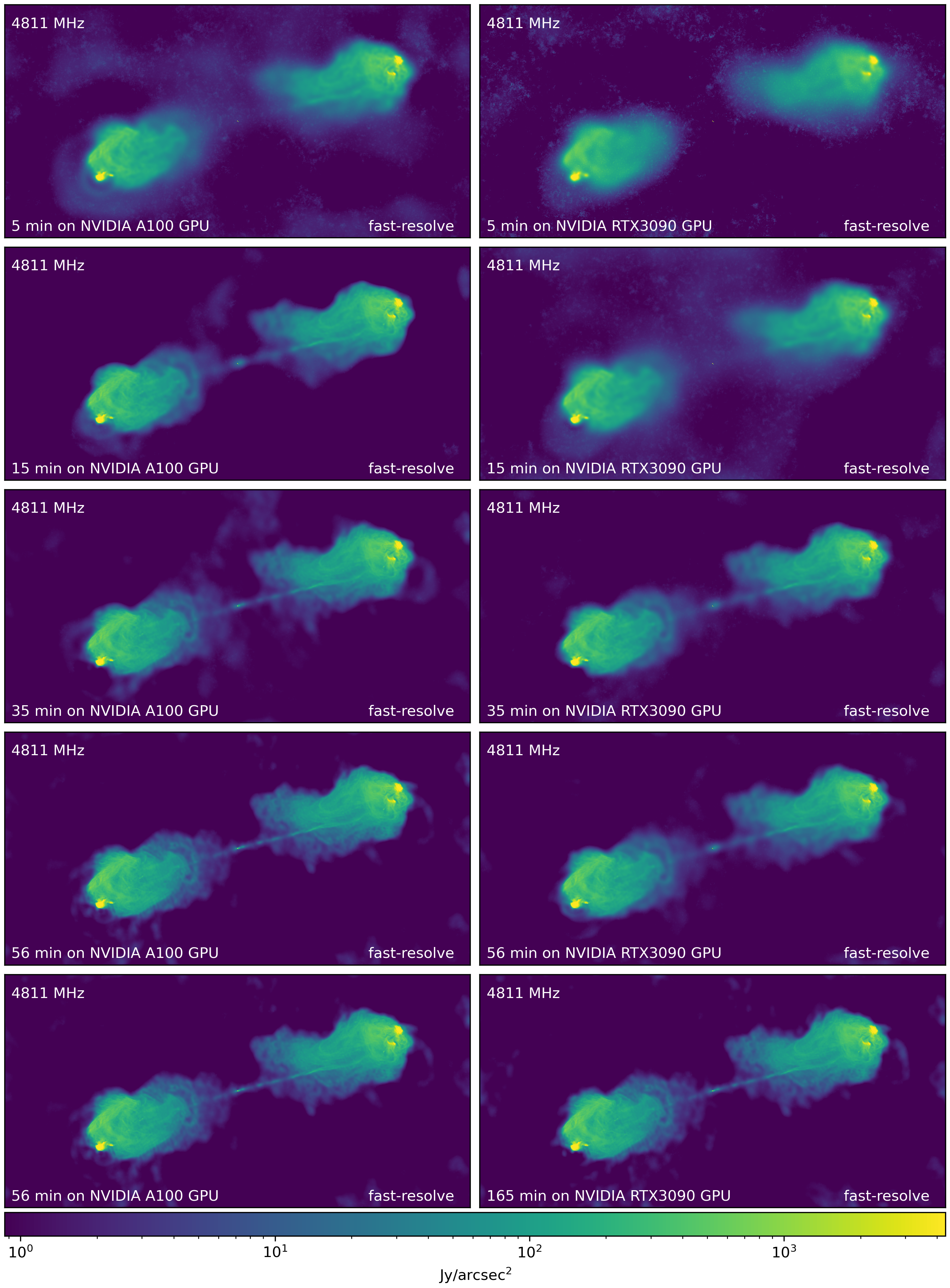}
         \caption{Same as Fig. \ref{fig:fast_gpu_a_VS_r2} but for the $4811\,\text{MHz}$ data. On an NVIDIA A100 GPU \texttt{fast-resolve} is $2-3$ times faster than on an NVIDIA RTX 3090 GPU. The same number of major and minor iterations are performed on both GPUs. Since the reconstruction on the A100 GPU is finished after $56\, \text{min}$, the bottom left panel is identical to the row above.}
         \label{fig:fast_gpu_a_VS_r4}
 \end{figure*}

The same comparison as in Fig. \ref{fig:fast_gpu_a_VS_r2} but for the C-band data is displayed in Fig. \ref{fig:fast_gpu_a_VS_r4}. As indicated in Tab. \ref{tab:obscyg}, the grid size we used for the C-band sky map is a factor of two larger along both spatial axes. Thus, in total, we have four times more pixels, increasing the computational cost of the algorithm. On the A100 GPU, the \texttt{fast-resolve} reconstruction was finished after $56\,\text{min}$, while on the RTX 3090 GPU, the reconstruction took $165\,\text{min}$. We believe that the larger difference between the two GPUs for the C-band data compared to the S-band data might be because, for the smaller grid size of the S-band reconstruction, the A100 GPUs were not fully utilized.

\begin{figure*}
    \centering
       \includegraphics[width=17cm]{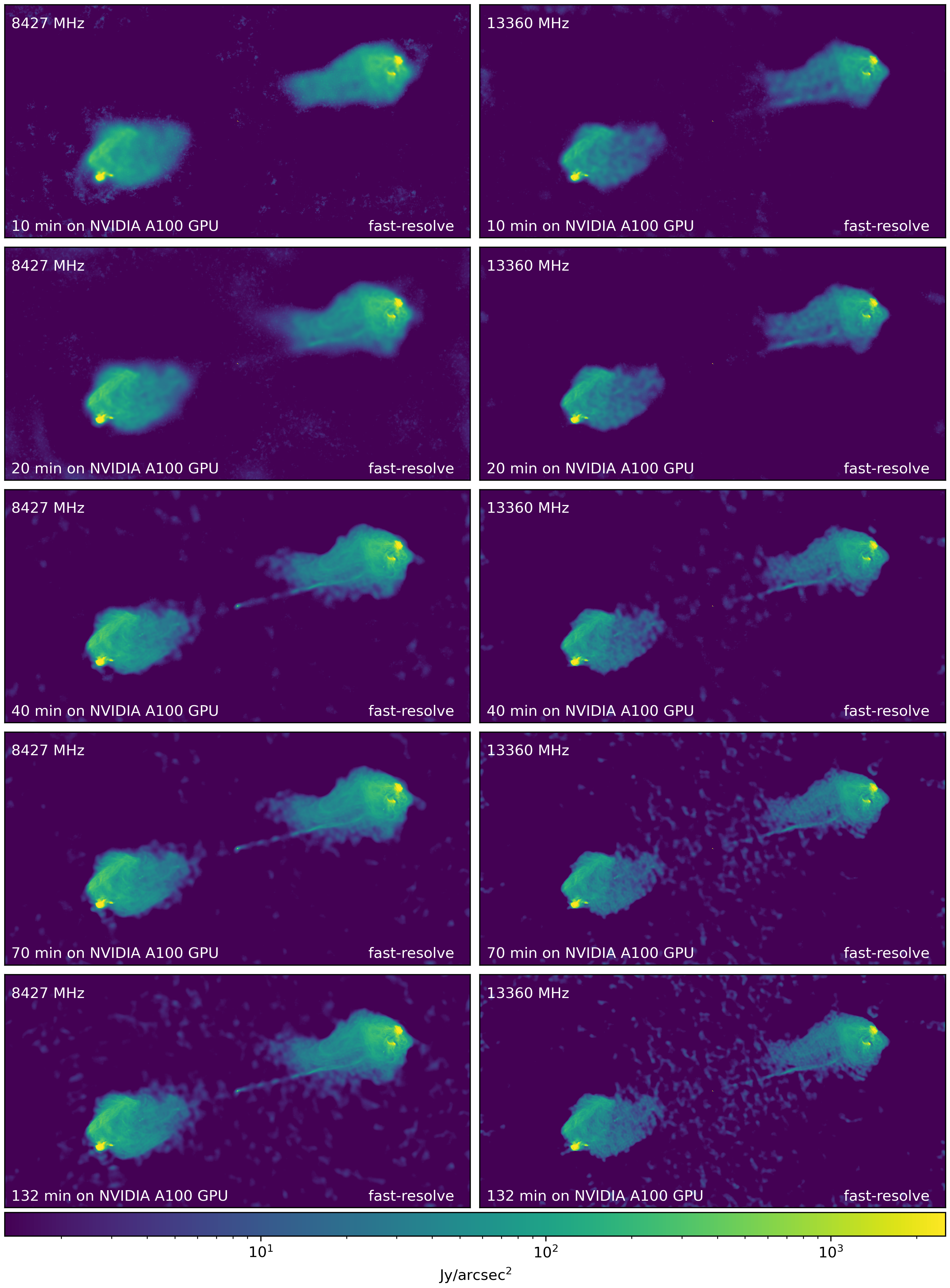}
         \caption{Snapshots of the \texttt{fast-resolve} Cygnus~A reconstruction of the $8427\,\text{MHz}$ and $13360\,\text{MHz}$ data on a NIFTy A100 GPU. The wall time after which the snapshot was taken is indicated in the bottom left of each panel.}
         \label{fig:x_ku_rct_gpu}
 \end{figure*}

For completeness, we also display snapshots of the X and Ku-band reconstructions in Fig. \ref{fig:x_ku_rct_gpu}. These reconstructions were only carried out on the A100 GPU. After $132\,\text{min}$ both reconstructions were finished.


\subsection{Application to MeerKAT data}\label{sec:applyspeed}
In this section, we present an application of \texttt{fast-resolve} to an L-band ($856-1712\,\text{MHz}$) MeerKAT \citep{2016mks..confE...1J} observation of the radio galaxy ESO~137-006.
Originally, this observation was presented in \cite{Ramatsoku20}.
The observation utilized all 64 MeerKAT antennas and the 4k mode of the SKARAB correlator.
The total on target observation time is $14\,\text{h}$ in full polarization with $4096$ channels.
For the VLA Cygnus~A observations above, only a single frequency channel was used for each band. Here, we use two sub-bands with each about $200$ channels (after averaging) and a bandwidth of approximately  $200\,\text{MHz}$.
Consequently the MeerKAT datasets are more than $400$ times larger than the VLA datasets of the previous section, making an evaluation of the radio interferometer response (Eq. \ref{eq:rime}) significantly more expensive.
A \texttt{resolve} reconstruction, where the exact response needs to be computed for each evaluation of the likelihood (Eq. \ref{sec:rvelh}), is computationally unfeasible for such a dataset.

\cite{Ramatsoku20} detected hitherto unknown collimated synchrotron threads linking the lobes of the radio galaxy ESO~137-006 in this observation. Since then, this data set has also been used by \cite{Dabbech22} for demonstrating a sparsity-based imaging algorithm. \cite{Dabbech22} also published the results as \texttt{FITS} files, allowing us to use them as a reference for validating the \texttt{fast-resolve} algorithm for a MeerKAT-sized reconstruction.

Technical details relating to the initial flagging and transfer calibration of the ESO~137-006 data using the CARACAL pipeline \citep{jozsa20} are given in \cite{Ramatsoku20}.
As in \cite{Ramatsoku20}, the data is averaged from 4096 to 1024 frequency channels and split into two sub-bands spanning $961-1145\,\text{MHz}$ and $1295-1503\,\text{MHz}$ that are relative free from radio frequency intereference, known as the LO and HI bands respectively. These subbands are then phase self-calibrated using WSClean multi-scale CLEAN \citep{Offringa2017} for imaging and CubiCal \citep{Kenyon18} for calibration.
We then image the data independently in the LO and HI band with \texttt{fast-resolve}.
Computational relevant details of the calibrated data in the two bands are listed in Tab. \ref{tab:obseso}.

\begin{table}
    \caption{MeerKAT ESO~137-006 observations. For both frequency bands the same field of view (Fov) is imaged.}
    \label{tab:obseso}
    \centering
    \begin{tabular}{c c c c c}
    \hline\hline
    Freq [MHz]& $N_{rows}$ & $N_{chan}$ & image size & Fov [deg]\\
    \hline
    961-1145  & 4598424 & 220 & $3600\times3600$ & $2\times2$ \\
    1295-1503 & 4598424 & 248 & $3600\times3600$ & $2\times2$ \\
    \hline
    \end{tabular}
\end{table}

\subsubsection{\texttt{fast-resolve} prior model}
As for the Cygnus~A observation, we built our prior model around the exponentiated Gaussian process model described in Sec. \ref{sec:applycygaintro} and in detail in \cite{Arras2021}. Nevertheless, for the MeerKAT observation, there is, besides the main source, ESO~137-006, also a second galaxy, ESO~137-007, in the field of view.
Due to the nature of the generative prior models in \texttt{resolve}, the full prior model can easily be composed of multiple components.
We model each of these sources with a separate exponentiated Gaussian process to decouple the prior models. Furthermore, due to the high sensitivity of MeerKAT, many compact background sources are detected. For the Cygnus~A reconstruction, we placed two point source models at the locations of the two point sources in the core of Cygnus~A. However, for the MeerKAT observation, the number of background sources is far too large to manually place point source models at their locations. Therefore, we use a third Gaussian process model to represent all the background sources outside of the models for ESO~137-006 and ESO~137-007. A sketch of the layout of the three models is shown in Fig. \ref{fig:fov}. The exact parameters for all Gaussian process models are listed in appendix \ref{app:parameso}.

\begin{figure}
    \resizebox{\hsize}{!}{\includegraphics{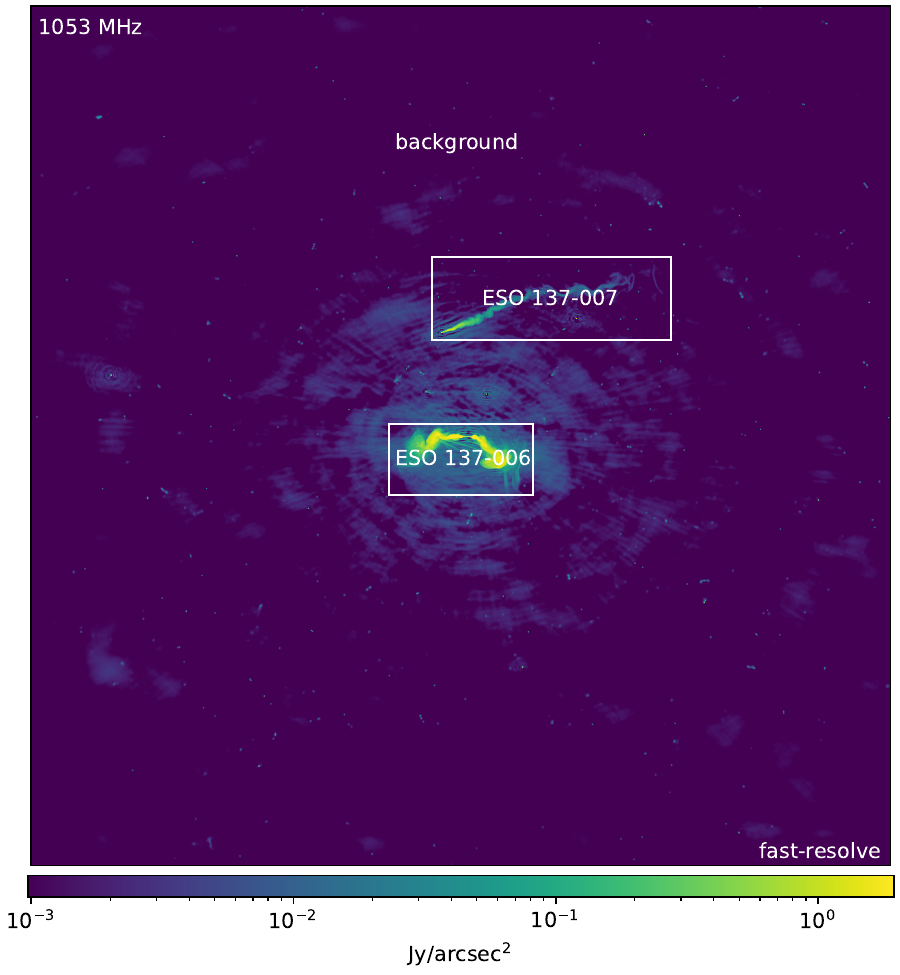}}
    \caption{Reconstruction of the ESO~137-006 MeerKAT observation in the LO subband ($961-1145\,\text{MHz}$) with \texttt{fast-resolve}. The radio galaxies ESO~137-006 and ESO~137-007, as well as the background sources, are modeled with independent Gaussian processes. White boxes indicate the spatial extent of each Gaussian process. Zooms on the reconstructions of ESO~137-006 and ESO~137-007 for both subbands are shown in Fig. \ref{fig:eso006}, \ref{fig:eso007}, \ref{fig:eso006_high}, and \ref{fig:eso007_high}.}
    \label{fig:fov}
 \end{figure}

\subsubsection{Imaging of \cite{Dabbech22}} \label{sec:dabbechcalimag}
In \cite{Dabbech22}, ESO~137-006 is imaged with WSClean \citep{Offringa2014,Offringa2017} and two convex optimization algorithms in the same bands $961-1145\,\text{MHz}$ and $1295-1503\,\text{MHz}$ using the same data as we do.
The two convex optimization algorithms of \cite{Dabbech22} utilize two different regularizers. One regularizer is named uSARA, promoting sparsity in a wavelet basis. The other regularizer, AIRI, originally presented in \cite{Terris23}, is based on a neural network denoiser. The resulting images for both regularizers are compared with results from WSClean, showing significant improvements in the image quality. The exact computational costs of all three imaging algorithms are reported in \cite{Dabbech22}. Imaging with the sparsity promoting regularizer, uSARA needed about $1500$ to $3000$ CPU hours per band to converge. With the neural network-based denoiser AIRI, around $900$ to $1600$ CPU hours plus around $5$ GPU hours where needed.

\subsubsection{Imaging results}
\begin{figure*}
    \centering
       \includegraphics[width=17cm]{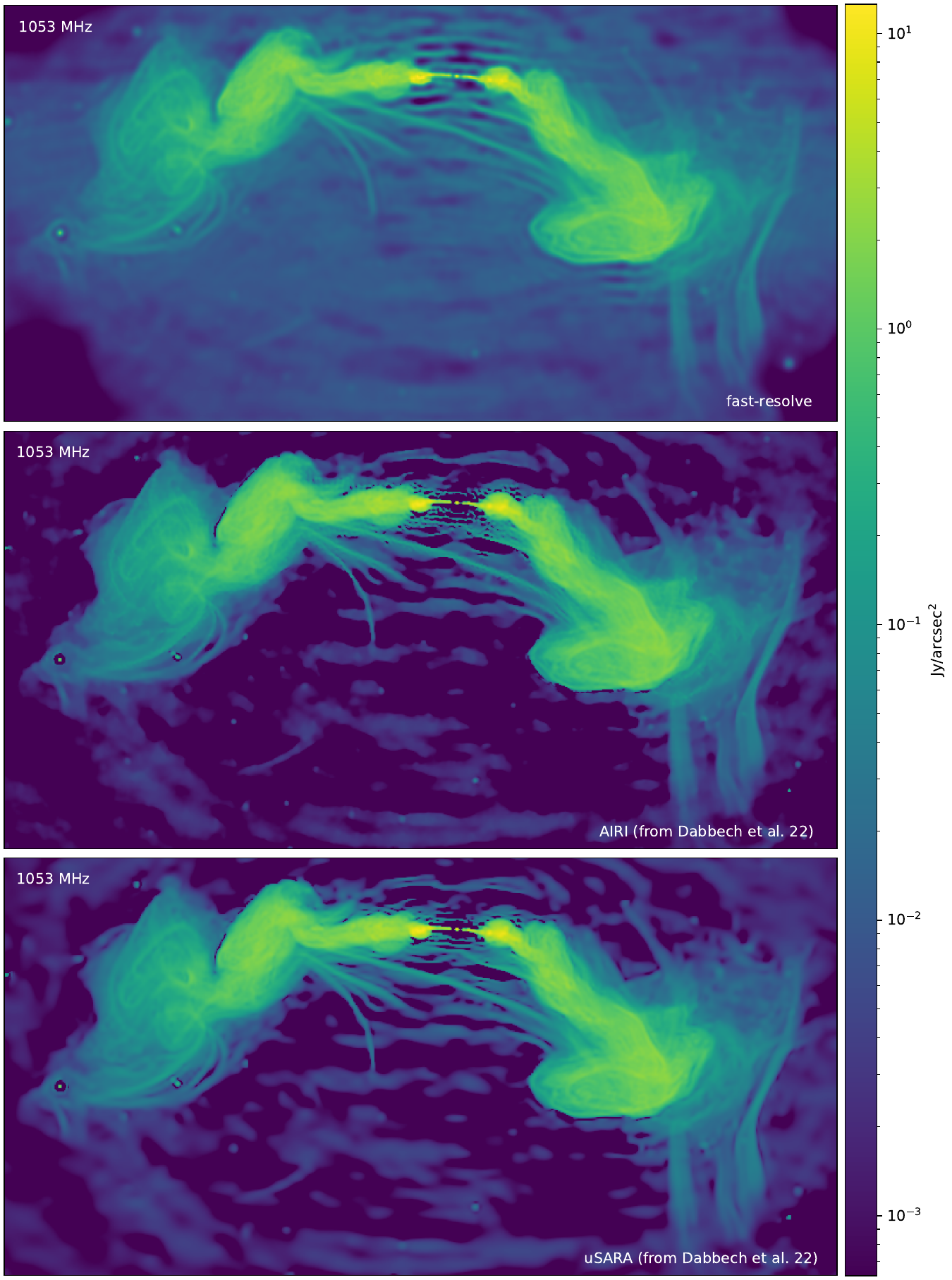}
         \caption{Radio Galaxy ESO~137-006 in the phase center of the observation. The top panel shows the \texttt{fast-resolve} reconstruction. For comparison the lower two panels display the reconstructions form \cite{Dabbech22} with the AIRI and uSARA regularizers. All reconstructions, but especially \texttt{fast-resovle}, are affected by the suboptimal calibration.}
         \label{fig:eso006}
 \end{figure*}

\begin{figure*}
    \centering
       \includegraphics[width=17cm]{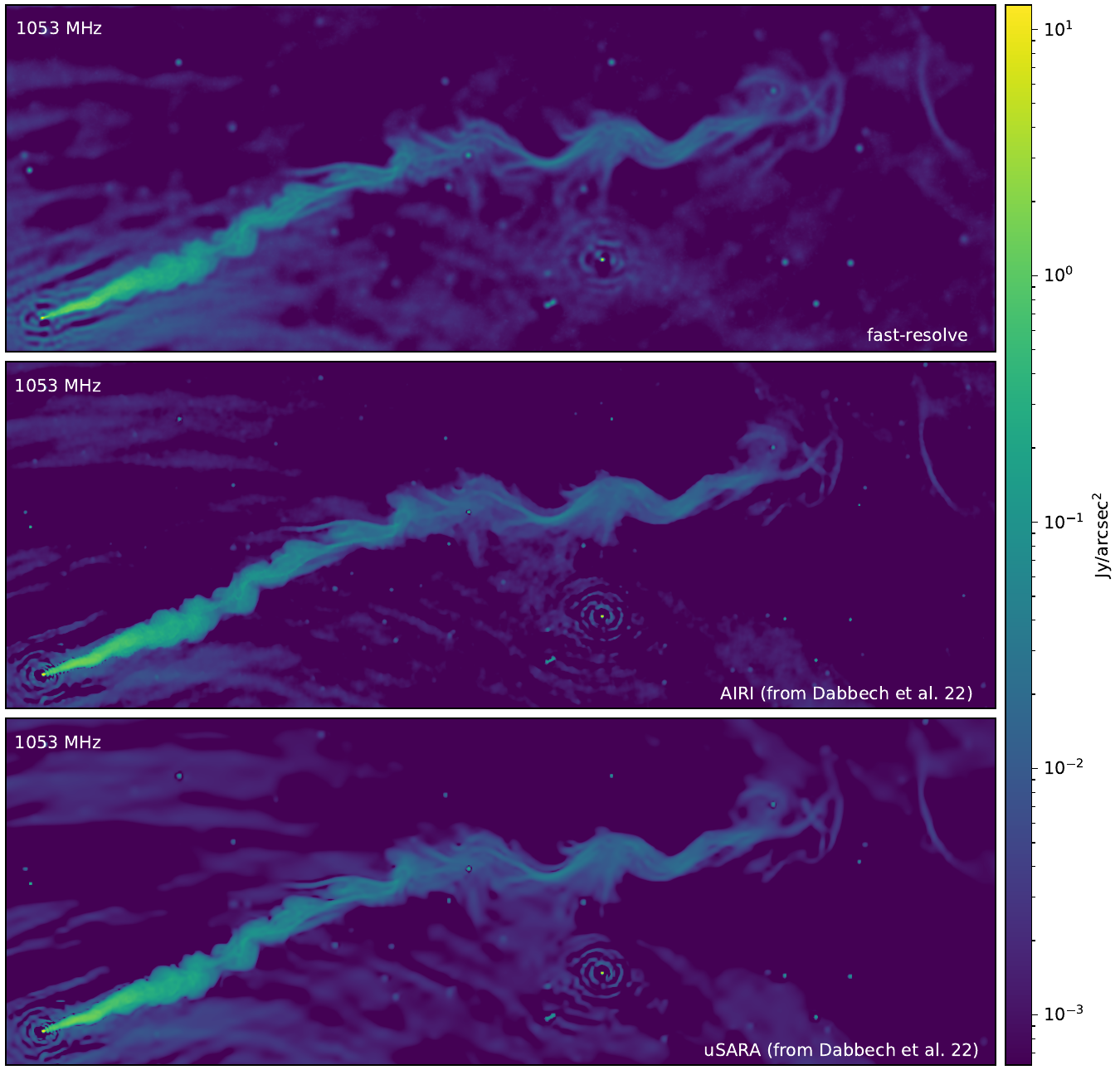}
         \caption{Radio Galaxy ESO~137-007 north of the phase center of the observation. The top panel shows the \texttt{fast-resolve} reconstruction. For comparison the lower two panels display the reconstructions form \cite{Dabbech22} with the AIRI and uSARA regularizers. All reconstructions, but especially \texttt{fast-resovle}, are affected by the suboptimal calibration.}
         \label{fig:eso007}
 \end{figure*}

\begin{figure*}
   \centering
      \includegraphics[width=17cm]{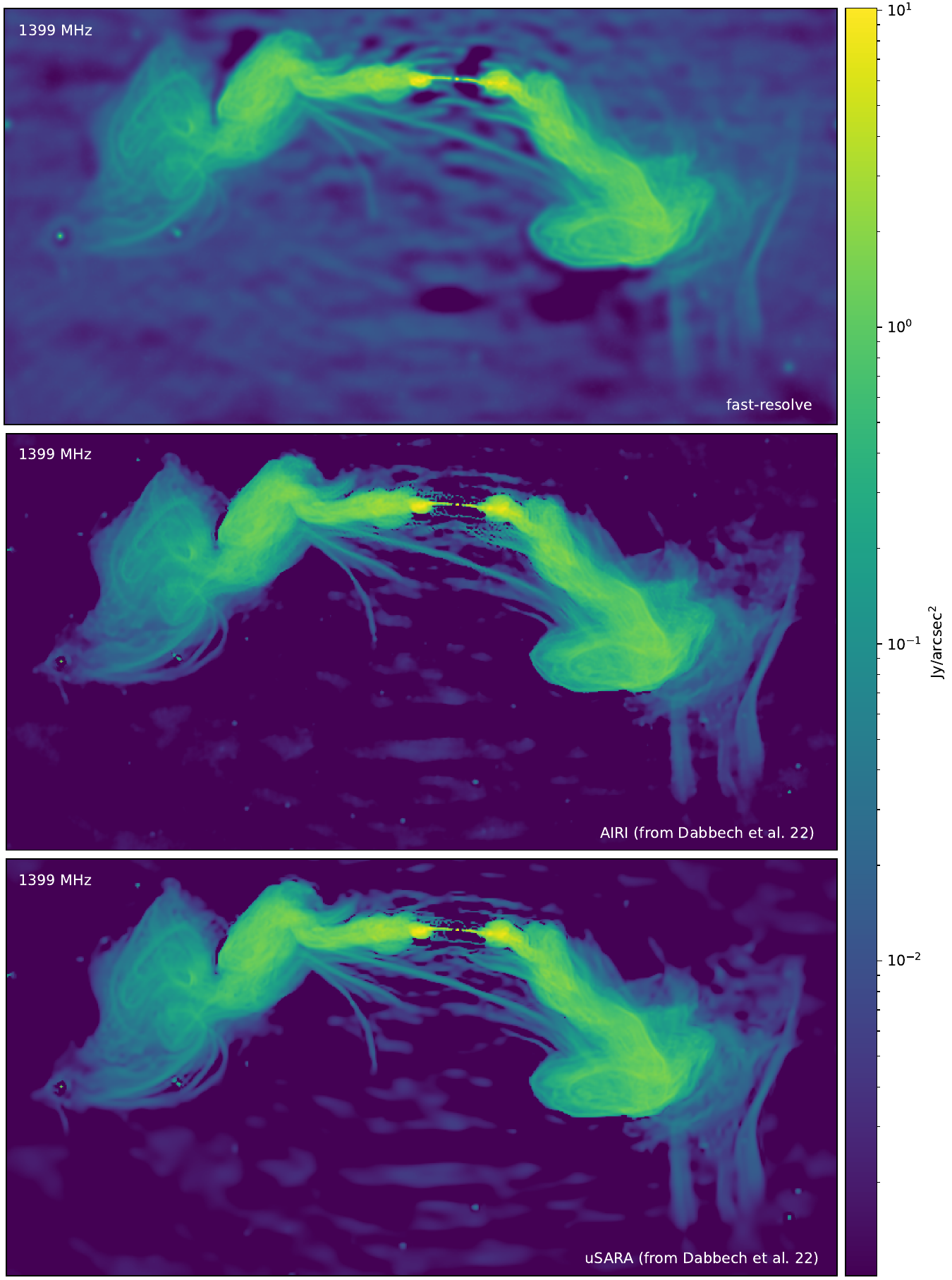}
        \caption{Like Fig. \ref{fig:eso006} but at $1399\,\text{MHz}$.}
        \label{fig:eso006_high}
\end{figure*}

\begin{figure*}
   \centering
      \includegraphics[width=17cm]{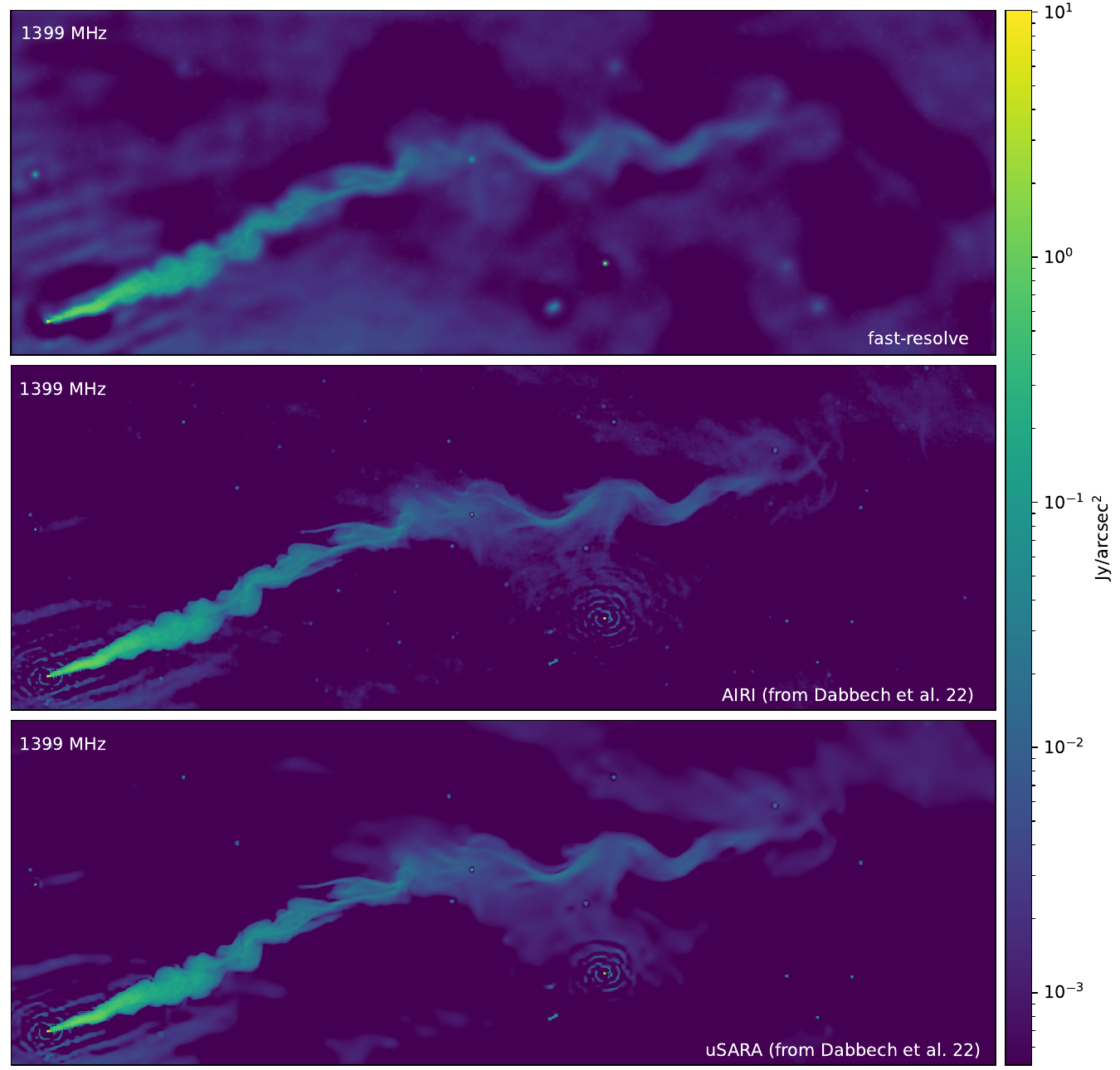}
        \caption{Like Fig. \ref{fig:eso007} but at $1399\,\text{MHz}$.}
        \label{fig:eso007_high}
\end{figure*}

Fig. \ref{fig:fov} depicts the results of the \texttt{fast-resolve} reconstruction for the LO band data. Zooms on the two radio galaxies are shown in Fig. \ref{fig:eso006} and Fig. \ref{fig:eso007} for the LO band and in Fig. \ref{fig:eso006_high} and Fig. \ref{fig:eso007_high} for the HI band. Thereby, Fig. \ref{fig:eso006} and Fig. \ref{fig:eso006_high} display the ESO~137-006 radio galaxy while Fig. \ref{fig:eso007} and \ref{fig:eso007_high} show the radio galaxy ESO~137-007 north of it. These figures also include the convex optimization reconstruction results of \cite{Dabbech22} for comparison and validation. The imaging results of \texttt{fast-resolve} and \cite{Dabbech22} are consistent. The \texttt{fast-resolve} maps have higher background artifacts than the convex optimization maps.
In \cite{Ramatsoku20}, collimated synchrotron threads between the two lobes of the galaxy ESO~137-006 were detected.
Besides these already known threads, the \texttt{fast-resolve} map shows additional threads north of the core of the galaxy. These additional threads are probably artifacts in the image.
With a lower intensity, similar features are also found in the maps of \cite{Dabbech22} and are mostly considered artifacts.

We believe these features are imaging artifacts and originate from suboptimal calibration.
Due to the very flexible
prior model, both \texttt{fast-resolve} and \texttt{resolve} are very sensitive the data.
In the case of suboptimal calibration, this leads to imaging artifacts.
At present, self-calibration with \texttt{fast-resolve} is not yet possible.
For the future, the integration of a self-calibration routine into \texttt{fast-resolve} is planned, which could potentially mitigate such artifacts.

Additionally, a dedicated prior model for compact sources could improve the \texttt{fast-resolve} reconstructions. At present, point sources can either be modeled by manually placing inverse gamma distributed pixels at their locations (see Sec. \ref{sec:rvepr}) or by including them in the Gaussian process model.
For the VLA Cygnus~A reconstruction, the point sources in the nucleus were modeled by manually placing inverse gamma distributed pixels at their locations.
Thereby, imaging artifacts around the two-point source could be avoided, despite them being very bright.
Due to the high number of background sources, this is impractical for the ESO~137-006 reconstruction.
Compact sources are therefore also modelled using an exponentiated Gaussian process prior.
A dedicated prior model that is also applicable to observations with a very high number of background sources could improve their reconstruction.

\subsubsection{Computational costs}
Despite the high number of visibilities and frequency channels, the computational costs of \texttt{fast-resolve} are still moderate since \texttt{fast-resolve} only needs to evaluate the full radio interferometric measurement equation (Eq. \ref{eq:rime}) in the major cycles. The \texttt{fast-resolve} reconstructions of both frequency bands were done on an NVIDIA A100 GPU. Before the actual \texttt{fast-resolve} imaging, the convolution kernels for the response and the noise were constructed on a CPU.
For both frequency bands, less than two hours were needed on a single core of an Intel Xeon CPU to compute the convolution kernel.
For imaging, 24 hours using an NVIDIA A100 GPU and eight cores of an Intel Xeon CPU were needed. Since most of the \texttt{fast-resovle} operators, except for the major update steps, are done on the GPU, the CPU was mostly idle during imaging.
In comparison, the computational costs of the \cite{Dabbech22} reconstructions are significantly higher as they range between $900$ and $3000$ CPU hours per band and algorithm. \cite{Dabbech22} for comparison also presented a WSClean reconstruction. For the WSClean reconstruction \cite{Dabbech22} report a total computational cost of $132$ and $236$ CPU hours for the two imaging bands.
Although CPU and GPU hours are not directly comparable, this underlines the computational performance and applicability of \texttt{fast-resolve} to imaging setups with massive datasets.


\section{Conclusion}
This paper introduces the fast Bayesian imaging algorithm \texttt{fast-resolve}.
\texttt{fast-resolve} combines the accuracy of the Bayesian imaging framework \texttt{resovle} with computational shortcuts of the \texttt{CLEAN} algorithm.
This significantly broadens the applicability of Bayesian radio interferometric imaging.
\texttt{fast-resolve} transforms the likelihood of the Bayesian radio interferometric imaging problem into a likelihood of a deconvolution problem, which is much faster to evaluate.
Using the major/minor cycle scheme of \texttt{CLEAN}, \texttt{fast-resolve} corrects for inaccuracies of the transformed likelihood and accounts for the w-effect.
The accuracy of \texttt{fast-resolve} is validated on Cygnus~A VLA data by comparing with previous \texttt{resolve} and multi-scale \texttt{CLEAN} reconstructions.
The comparison shows that \texttt{fast-resolve} achieves the same resolution as \texttt{resolve}.
Likewise, the imaging artifacts are comparable and on a very low level.

Furthermore, the computational speed of \texttt{fast-resolve} is analyzed and compared to \texttt{resolve}, showing significant speedups for the VLA Cygnus~A data.
As \texttt{fast-resolve} is implemented in JAX, it can also be executed on a GPU, accelerating the reconstruction compared to the CPU by more than an order of magnitude.
For the single channel Cygnus~A VLA dataset, \texttt{fast-resolve} is at least $140$ times faster than \texttt{resolve} when executed on a GPU.
For datasets with more frequency channels, the computational advantages of \texttt{fast-resolve} can be even larger.

Additionally, we present a Bayesian image reconstruction of the radio galaxies ESO~137-006 and ESO~137-007 from MeerKAT data with \texttt{fast-resolve} and compare the results for validation with \cite{Dabbech22}.
The MeerKAT dataset is significantly larger than the VLA datasets, but the computational costs of \texttt{fast-resolve} remain moderate due to the major/minor cycle scheme.
A reconstruction of these sources with the classic \texttt{resolve} algorithm using the same amount of data would be computationally out of scope.
To the best of our knowledge, no other Bayesian radio interferometric imaging algorithm has been successfully applied to a dataset of similar size before.

\section{Data Availability}
The raw data of the Cygnus~A observation is publicly available in the NRAO Data Archive\footnote{\url{https://data.nrao.edu/portal/}} under project ID 14B-336.
The raw data for the ESO137-006 observation is publicly available via the SARAO archive\footnote{\url{https://archive.sarao.ac.za}} (project ID SCI-20190418-SM-01).
The \texttt{fast-resolve} reconstruction results are archived on zenodo\footnote{\url{https://doi.org/10.5281/zenodo.11549302}}.
The implementation of the \texttt{fast-resolve} algorithm will be integrated into the \texttt{resolve} algorithm\footnote{\url{https://gitlab.mpcdf.mpg.de/ift/resolve}}.

\begin{acknowledgements}
The MeerKAT telescope is operated by the South African Radio Astronomy Observatory, which is a facility of the National Research Foundation, an agency of the Department of Science and Innovation.
J. R. acknowledges financial support from the German Federal Ministry of Education and Research (BMBF) under grant 05A23WO1 (Verbundprojekt D-MeerKAT III). P. F. acknowledges funding through the German Federal Ministry of Education and Research for the project “ErUM-IFT: Informationsfeldtheorie für Experimente an Großforschungsanlagen” (Förderkennzeichen: 05D23EO1).
O. M. S.'s research is supported by the South African Research Chairs Initiative of the Department of Science and Technology and National Research Foundation (grant No. 81737).
\end{acknowledgements}

\bibliographystyle{aa}
\bibliography{references}

\begin{appendix}
\section{Prior parameters for Cygnus~A imaging} \label{app:paramcyga}
In Tab. \ref{tab:cygparams} we list the hyper parameters of the Gaussian process model for the diffuse emission in the Cygnus~A reconstructions. The exact definition of these parameters is explained in detail in \cite{Arras2021}.

\begin{table}
    \caption{Prior parameters for Cygnus~A reconstructions}
    \label{tab:cygparams}
    \centering
    \begin{tabular}{l c c c c}
    \hline\hline
    Cygnus~A, band: & S & C & X & Ku\\
    \hline
    zero mode offset      & 18  & 21  & 20  & 20  \\
    zero mode mean        & 1   & 1   & 1   & 1   \\
    zero mode stddev      & 0.1 & 0.1 & 0.1 & 0.1 \\
    fluctuations mean     & 5   & 5   & 5   & 5   \\
    fluctuations stddev   & 1   & 1   & 1   & 1   \\
    loglogavgslope mean   & -2  & -2  & -2.2& -2.2\\
    loglogavgslope stddev & 0.2 & 0.2 & 0.2 & 0.2 \\
    flexibility mean      & 1.2 & 1.2 & 1.2 & 1.2 \\
    flexibility stddev    & 0.4 & 0.4 & 0.4 & 0.4 \\
    asperity mean         & 0.2 & 0.2 & 0.2 & 0.2 \\
    asperity stddev       & 0.2 & 0.2 & 0.2 & 0.2 \\
    \hline
    \end{tabular}
\end{table}

\section{Prior parameters for ESO~137 imaging} \label{app:parameso}
In Tab. \ref{tab:esoparams_low} we list the hyper parameters of the Gaussian process model for the diffuse emission in the ESO~137 reconstructions. The exact definition of these parameters is explained in detail in \cite{Arras2021}.

\begin{table}
    \caption{Prior parameters for ESO~137 reconstructions. For the two frequency bands, the LO ( $961-1145\,\text{MHz}$) and the HI ( $1295-1503\,\text{MHz}$) band we used the same hyper parameters.}
    \label{tab:esoparams_low}
    \centering
    \begin{tabular}{l c c c c}
    \hline\hline
    ESO 137 & 006 & 007 & background\\
    \hline
    zero mode offset      & 13  & 13  & 13  \\
    zero mode mean        & 1   & 1   & 1   \\
    zero mode stddev      & 0.1 & 0.1 & 0.1 \\
    fluctuations mean     & 3   & 4   & 4   \\
    fluctuations stddev   & 1   & 1   & 1   \\
    loglogavgslope mean   & -2.4  & -2.4  & -2.0\\
    loglogavgslope stddev & 0.1 & 0.1 & 0.1 \\
    flexibility mean      & 0.1 & 0.1 & 1.2 \\
    flexibility stddev    & 0.01 & 0.1 & 0.4 \\
    asperity mean         & 0.01 & 0.01 & 0.2 \\
    asperity stddev       & 0.001 & 0.001 & 0.2 \\
    \hline
    \end{tabular}
\end{table}

\end{appendix}

\end{document}